\documentclass[10pt,twocolumn,journal]{IEEEtran}

\usepackage{graphicx}
\usepackage{tabularx}
\usepackage{times}
\usepackage{amsmath}
\usepackage{amsthm}
\newtheorem{theorem}{Theorem}
\newtheorem{lemma}{Lemma}
\usepackage{pifont}

\usepackage{amsfonts}
\usepackage{amssymb}
\usepackage{cite}
\usepackage{epsfig}
\usepackage{subfigure}
\usepackage[ruled]{algorithm2e}

\begin{document}
\title{Joint Optimization of Scheduling and \\Power Control in Wireless Networks: \\Multi-Dimensional Modeling and Decomposition}

\author{Lu~Liu,~\IEEEmembership{Student~Member,~IEEE,}
        Yu~Cheng,~\IEEEmembership{Senior~Member,~IEEE,}
        Xianghui~Cao,~\IEEEmembership{Senior~Member,~IEEE,}
        Sheng~Zhou,~\IEEEmembership{Member,~IEEE,}
        Zhisheng~Niu,~\IEEEmembership{Fellow,~IEEE}
        and~Ping~Wang,~\IEEEmembership{Senior~Member,~IEEE}
\thanks{Lu Liu and Yu Cheng are with the Department
of Electrical and Computer Engineering, Illinois Institute of Technology, Chicago,
IL 60616.
E-mail: lliu41@hawk.iit.edu.}
\thanks{Xianghui Cao is with School of Automation, Southeast University, Nanjing 210096, China.}
\thanks{Sheng Zhou and Zhisheng Niu are with Tsinghua National Laboratory for Information Science and Technology, Tsinghua University, Beijing 100084, China.}
\thanks{Ping Wang is with School of Computer Engineering, Nanyang Technological University, Singapore 639798.}
}

\maketitle

\begin{abstract}
The energy efficiency of future networks is becoming a significant and urgent issue, calling for greener network designs. At the same time, rapid development of wireless networks shows a trend of increasing complexity in network structure and resource space, leading to that optimizing the energy efficiency of such networks requires a joint solution over multi-dimensional resource space. However, the coupled resource dimensions and growing problem scales bring great challenges in obtaining the optimal solutions. In this paper, we develop a multi-dimensional network model on the basis of tuple-links associated with transmission patterns (TPs) and formulate the optimization problem as a TP based scheduling problem which jointly solves transmission scheduling, routing, power control, radio and channel assignment. In order to tackle the complexity issues raised from coupled resource dimensions, we propose a novel algorithm that decomposes the coupling scheduling and power control by exploiting the delay column generation technique to recursively solve a master problem for scheduling and a sub-problem for power allocation. Further, we theoretically prove that the performance gap between the proposed algorithm and the optimum is upper bounded by that for the sub-problem solution, where the latter is derived by solving a relaxed version of the sub-problem. Numerical results demonstrate the effectiveness of the multi-dimensional framework and the benefit of the proposed joint optimization in improving network energy efficiency.
\end{abstract}


\IEEEpeerreviewmaketitle

\section{Introduction}\label{sec:introduction}

Energy efficiency of next generation wireless networks is a critical and urgent issue. The key to improving energy efficiency of wireless networks relies on configuring and allocating various network resources in temporal, spatial, spectral and power dimensions in terms of routing, link scheduling, channel allocation and power control. Generally, different network resources are coupled such that they cannot be determined independently for optimal performance, which demands a joint optimization solution. What's more, in order to meet the rapidly growing traffic demands, wireless networks are evolving into more and more complex structures and hence large scales of joint optimization problems. Obtaining a joint optimization solution over wireless networks becomes a challenging issue, which motivated us to develop more efficient solutions.

Many wireless networks can be abstracted as that each network node has multiple radio interfaces operating on multiple available wireless channels, yielding the generic multi-radio multi-channel (MR-MC) network model with multi-dimensional resource space  \cite{naeem2014resource,ho2014harnessing,cheng2}. With this model for the joint optimization problem, the optimization variable can be viewed as a compound of multiple resource allocation strategies, including selection of transmitters and receivers for transmission links, radio and channel assignment, transmit power control, routing and link scheduling.

The existing studies on energy-efficient networking in MR-MC networks have addressed the joint optimization issues over different dimensions, but a generic joint optimization solution over the whole multi-dimensional space (especially when power control is involved) is still not available, to the best of our knowledge.
Radio/channel assignment and transmission scheduling in MR-MC networks have been studied with the objective to maximize network capacity \cite{chen2011game,saifullah2014distributed,chengsystematic,cheng1}. Specifically, protocol interference model is widely adopted to characterize the interferences among links as a conflict graph, over which independent set based scheduling is then used to facilitate a linear programming (LP) based formulation \cite{li2015optimal,cheng2014systematic,cao2016distributed}. However, such a model simplifies transmission links to be either deactivated or activated with fixed transmit power, which can neither model dynamic power assignment nor accurately reflect the practical interference magnitude. The more realistic signal-to-interference-plus-noise ratio (SINR) based physical interference model can model transmission interferences under the power control. Link scheduling for capacity optimization under the physical interference model has been studied in \cite{wan2011wireless,zhou2012distributed,le2010longest}, but is limited to single-channel scenarios. How to incorporate physical interference model based power control into multi-dimensional resource space so as to provide energy-efficient joint scheduling and power control solution remains a challenging issue.

To this end, we apply the tuple-link based multi-dimensional network model \cite{chengsystematic}, with which the joint allocation over multi-dimensional resource space is reduced to the scheduling of tuple-links.
Further, we propose a new concept of \textit{transmission pattern} (TP) which integrates both scheduling and power control to facilitate LP formulation of the joint optimization problem. A TP is defined as a vector of transmit power assignment of all the tuple-links in the network. In a TP, the SINR at the receiver of each tuple-link can be calculated based on the power allocation of all the tuple-links so that the transmission capacity of the tuple-link can be determined according to the Shannon-Hartley equation. Therefore, a TP characterizes a possible transmission state in the network, including the resource allocation information across all the dimensions. By considering discretized transmit power levels, the joint scheduling and power assignment problem in the multi-dimensional resource space can be ultimately transformed into a scheduling problem of a finite number of TPs, which facilitates an LP formulation. This TP based scheduling problem is formulated in a similar manner as that of independent set based scheduling \cite{cheng2,chengsystematic}, but is compatible with physical interference model and flexible power allocation. The solution to the TP based optimization provides joint scheduling and power control, as well as the resource allocation on all the other resource dimensions.

The TP based scheduling however leads to an extremely large problem scale due to exponentially growing number of TPs. We then propose a decomposition based approach by leveraging delay column generation (DCG), which starts with an initial subset of TPs and then gradually adding new TPs that can improve the objective value. The DCG based method repeatedly solves a master problem and a sub-problem, where the master problem performs scheduling on existing TPs and the sub-problem searches for a new entering TP by solving a maximum utility problem.
We further reveal that the sub-problem is indeed to find the most energy-efficient TP according to the information extracted from the existing TPs, and show that it is equivalent to find the optimal power allocation over tuple-links. Thus the joint optimization problem is decomposed into an iterative procedure combining scheduling phase and power control phase, while optimality remains intact during the decomposition.

As solving the sub-problem still incurs high computational complexity in searching over the entire TP space, we further propose a greedy algorithm to solve the sub-problem efficiently. Moreover, through theoretical analysis, we prove that the performance gap between the achieved and optimal solutions is upper bounded by the gap achieved in solving the sub-problem, which can be derived by solving a relaxed sub-problem.

Some preliminary results appearing in \cite{liu2016dafee} focus on the simple single-hop scenario  where per-link traffic demand is explicitly specified in the optimization formulation. This paper extends the proposed framework to generic multi-hop scenarios with multiple commodity flows. The obtained joint solution further provides routing information in a way that the source to destination paths for each commodity flow are implied by the obtained schedule of the links. In addition, a new modeling method and the concept of interference coefficient is introduced in this paper which seamlessly integrates radio conflict and co-channel interference. Further, in solving the sub-problem, a greedy algorithm is proposed in this paper instead of the learning based algorithm in \cite{liu2016dafee} for higher computational efficiency.

The main contributions of this paper are summarized as follows:

\begin{enumerate}
  \item We formulate a joint optimization framework for energy-efficient networking in the multi-dimensional resource space, and translate the original optimization into a TP based scheduling problem which is a linear program.
  \item To solve the large-scale TP scheduling problem, we develop a decomposition approach by exploiting the DCG technique that decomposes the optimization problem into scheduling (master problem) and energy-efficient TP selection (sub-problem). In addition, we propose a greedy algorithm to efficiently solve the sub-problem.
  \item We theoretically prove that the performance gap of the original problem's solution is bounded by that of the sub-problem, and derive the latter by formulating and solving a relaxed version of the sub-problem.
  \item We present numerical results to demonstrate the energy efficiency improvement of joint scheduling and power control, and analyze how the allocations of multi-dimensional resources affect the energy efficiency in the network.
\end{enumerate}

The remainder of this paper is organized as follows. Section 2 reviews more related work. Section 3 describes the system model and problem formulation. Section 4 presents the decomposition framework and algorithm, with the performance bound of the proposed algorithm analyzed in Section 5. Section 6 presents numerical results, and Section 7 gives the conclusion remarks.

\emph{Notations:} Throughout this paper, we use $|\mathcal{A}|$ to denote the size of set $\mathcal{A}$. Boldfaced capital letters are used to denote matrices, while boldfaced lower-case letters are used to denote vectors. All the vectors are column vectors by default, and the transpose of a matrix $\bold{A}$ is denoted as $\bold{A}'$.

\section{Related Work}

Energy-efficient networking has gained great attention in the literature, especially for networks with multi-dimensional resource space such as heterogeneous networks \cite{xie2012energy}, cognitive radio networks \cite{lu2013energy,joint2,wang2015joint} and networks with device-to-device communications \cite{zhou2014distributed,wang2013energy}.
Resource allocation for heterogeneous cognitive radio network is studied in \cite{xie2012energy}, where a Stackelberg game approach is adopted with gradient based iteration algorithm as a solution.
Channel assignment and power control are investigated in \cite{lu2013energy} which aims to maximize energy efficiency of cognitive radio networks and maps the optimization problem to a maximum matching problem.
Similarly, a joint solution of channel and power allocation is proposed in \cite{joint2}, with the objective of maximizing overall network throughput. In that paper, physical interference model is applied and the problem is solved by formulating a bargaining based cooperative game.
The work in \cite{wang2015joint} investigates the joint optimization of spectrum and energy efficiency in cognitive networks with power and subchannel allocation, where the authors propose a tradeoff metric based problem transformation and exploiting convex problem structure.
In \cite{zhou2014distributed}, an energy efficiency maximization problem is formulated as a non-convex program, which is then transformed into a convex optimization problem with nonlinear fractional programming.
The authors in \cite{wang2013energy} consider joint radio and power allocation for energy efficiency optimization, and develop an auction game based approach. The above works focus on specific network scenarios or configurations, which could not be applied to generic MR-MC networks with multi-dimensional resource spaces. Furthermore, as most of them focus on channel and power allocation, joint optimizations incorporating link scheduling has not been well studied.

In \cite{liu2014optimizing}, the problem of energy efficiency optimization in MR-MC networks is considered to derive radio/channel assignment and scheduling solutions for optimal energy efficiency under the requirement of full network capacity. A similar approach is adopted in \cite{liu2014energy} to minimize energy consumption with guaranteed capacity requirement. The problem is solved with a decomposed approach due to the large scale solution space. While these works take protocol interference model to simplify the scheduling problem, the more realistic physical interference model is applied in \cite{anderson2014optimization} for a joint scheduling and radio configuration problem. However, power control is not accounted, i.e., they all use fixed transmit power in the formulation.

To take power allocation into resource allocation, the authors in \cite{sheng2014utility} propose an algorithm to jointly allocate channel and power with a utility based learning method in a decentralized manner. The utility is characterized by the transmission rate achieved by links and the solution can maximize the sum rate of links, but without considering energy cost. For energy efficiency optimization, a joint cell selection and power allocation problem for heterogeneous networks is formulated in \cite{mensah2016energy} and solved with a Lagrange dual based method, where the proposed model does not apply to generic multi-dimensional resource space.
The work in \cite{zhao2016fundamental} proposes a two-step approach which first fixes transmit power to solve for scheduling and then optimizes the transmit power on the solved scheduling solution. However, such a decomposition will lead to sub-optimality since scheduling and power control are indeed solved separately. A joint solution of scheduling, channel allocation and power control is proposed in \cite{li2015energy}, but the achievable data rate on links is assumed constant. This model cannot fully reflect the link capability, since the latter is characterized by the real-time SINR at the receiver.
In sum, in the literature, a joint optimization solution towards energy efficient networking over the multi-dimensional resource space including routing, link scheduling, radio/channel assignment and power allocation has not been fully investigated, which is then to be studied in this paper.

\section{Problem Formulation}\label{sec:system}

\begin{table}[!t]
\renewcommand{\arraystretch}{1.3}
\caption{Notations}
\label{notation}
\centering
\begin{tabular}{c|c}
\hline
\hline
$\mathcal{N}$ & node set \\
\hline
$\mathcal{C}$ & set of channels \\
\hline
$\mathcal{R}$,$\mathcal{R}_v$ & radio set, radio set of node $v$\\
\hline
$\mathcal{P}$ & set of available power levels\\
\hline
$\lambda$& flow commodity\\
\hline
$q^{(\lambda)}$ & demand of commodity $\lambda$\\
\hline
$l$, $\mathcal{L}$ & tuple-link, set of all tuple-links\\
\hline
$\gamma_l$ & SINR of tuple-link $l$\\
\hline
$g_{ll'}$ & generalized interference coefficient between $l$ and $l'$\\
\hline
$\alpha$, $\mathcal{A}$ & TP, set of all TPs\\
\hline
$t_{\alpha}$ & portion of transmission time assigned to TP $\alpha$\\
\hline
$p_{l,\alpha}$, $r_{l,\alpha}$ & transmit power, capacity of link $l$ achieved in TP $\alpha$\\
\hline
$u,U,\tilde{u}$ & link utility, system utility, utopian utility\\
\hline
\hline
\end{tabular}
\end{table}

\subsection{Network Model}
Consider a generic MR-MC network with node set $\mathcal{N}$. Each node $v\in\mathcal{N}$ is equipped with one or multiple radio interfaces which are denoted as radio set $\mathcal{R}_v$. Define the set of all radios in the network as $\mathcal{R}$, thus $\mathcal{R}=\cup _{v\in \mathcal{N}}\mathcal{R}_v$. For each radio, all the other nodes' radios within its maximum transmission range are defined as its neighbors. For a non-isolated node, each of its radios can set up transmission links to all its neighbors. Denote the maximum transmit power of a radio as $p_{\text{max}}$, and assume that the transmit power of each radio takes value from a discrete set of power levels $\mathcal{P}$. There is a number of non-overlapping channels available to each radio. We denote all the channels as set $\mathcal{C}$.

The objective is to minimize the total energy consumption in the network under traffic demand requirement. Denote the set of multiple commodity flows as $\{1,\cdots ,\lambda, \cdots ,\Lambda \}$. Each flow $\lambda$ is specified by its corresponding source-destination node pair and flow demand. Therefore, it requires to jointly address: routing, link scheduling, radio and channel assignments, and transmit power control. In this optimization, the scheduling problem is to select transmission links and decide the transmission time for them. It can be seen that the joint optimization problem involves both continuous and discrete decision variables, making it a mixed-integer problem which is known of high complexity. In what follows, we present a tuple-link based framework to remodel the network, which facilitates an LP formulation and problem decomposition.

\begin{figure}[ht]
  \centering
  \scalebox{0.3}
  {\includegraphics{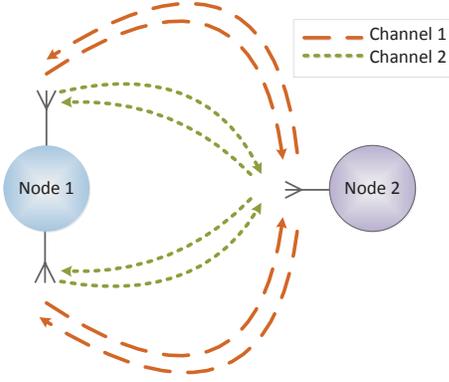}}
  \caption{Tuple-link example.}
  \label{fig:link}
\end{figure}

A {\em tuple-link} is defined as a combined resource allocation for a transmission indicating the transmitter radio, the receiver radio\footnote{Tuple-link is directional since the transmitter and receiver are specified.} and the operating channel \cite{cheng2}. Denote $\mathcal{L}$ as the set of all the tuple-links in the network. Tuple-link only exists when there exists a corresponding physical link (between a radio and its neighbor); a physical link can be mapped to multiple tuple-links. Fig. \ref{fig:link} gives an example of tuple-links between two nodes, where each node has two radios and 2 channels are available. As shown by the dash lines, there exist 8 tuple-links specified by different transmitters, receivers or channels.
With this tuple-link based framework, the above optimization problem becomes to jointly solve scheduling and power control of the tuple-links since radio and channel assignment is encapsulated into tuple-link selection. In the rest of this paper, we use ``link'' to stand for ``tuple-link'' unless stated otherwise.

In a wireless network, links may suffer from interference from other concurrent transmitting links. In this paper, we consider physical interference model, in which the capacity of a link can be characterized by the SINR at the receiver. For a link $l\in\mathcal{L}$, the received SINR is defined as
\begin{equation}\label{eq:SINR}
\gamma_{l}=\frac{g_{l}p_{l}}{I_{l}+\sigma^2}
=\frac{g_{l}p_{l}}{\sum\limits_{l' \in \mathcal{L}\setminus l}g_{l'l}p_{l'}+\sigma^2}
\end{equation}
where $g_{l}$, $p_{l}$, $I_{l}$, $\sigma^2$ denote the link gain, transmit power, received interference and the noise power, respectively. Particularly, $g_{l'l}$ is used to characterize the strength of interference from $l'$ to $l$ and will be discussed in details in the following. The capacity (maximum achievable transmission rate) of link $l$ can be expressed as
\begin{equation}\label{eq:a}
r_{l}=B_l \log_2(1+\gamma_{l})
\end{equation}
where $B_l$ is the corresponding channel bandwidth of $l$.

The link gain of $l$ is given as $g_{l}=\rho(d_{l})$, where $d_l$ is the distance between $l$'s transmitter and receiver and $\rho(\cdot)$ is a function of $d_l$ (e.g., the path loss function). Similarly, $g_{ll'}=\rho(d_{ll'})$ if $l$ and $l'$ transmit in the same channel with $d_{ll'}$ as the distance from $l$'s transmitter to $l'$'s receiver. Notice that two links on different channels will not generate co-channel interference to each other and in this case $g_{ll'}=0$.

Besides co-channel interference, two links may not work simultaneously due to radio conflict. For example, two links cannot share the same transmitter radio for simultaneous transmissions. The radio conflict is usually expressed as integer constraints in optimization problems \cite{li2015energy,anderson2014optimization} or considered separately aside from other resource allocation \cite{liu2016dafee}. In order to facilitate linear programming, we extend the definition of $g_{ll'}$ to cover the radio conflict relationship, and specifically redefine it as \textit{interference coefficient}. We apply a very large interference coefficient between links sharing the same radio. For example, if $l$ and $l'$ have the same transmitter radio, then we may set $g_{ll'}=\infty$. Assume all the radios in the network are half-duplex, and at any time a radio interface can be occupied by at most one link for transmission. In this case, the interference coefficient can be defined as
\begin{equation*}
  g_{ll'}\triangleq
  \begin{cases}
   \rho(d_{ll'}), & \text{if $l$ and $l'$ use different radios}\\ & \text{~~and are on the same channel;}\\
   0, & \text{if $l$ and $l'$ use different radios}\\ & \text{~~and are on different channels;}\\
  \infty,  & \text{if $l$ and $l'$ share one or two radios.}
  \end{cases}
\end{equation*}
where $\infty$ stands for a significantly large number. According to this definition, the SINR expression in Eq. (\ref{eq:SINR}) is able to characterize both the radio conflict and interference, based on which the optimization problem can be formulated without additional radio constraints.

The generalized radio conflict model can adapt to other network scenarios that have different types of radio constraints by correspondingly adjusting the values in the definition. For example, in CDMA networks, radios are allowed to receive from multiple transmitters at the same time. In this case, the interference coefficient between two links sharing the same receiver radio can be defined according to the distance function and coding gain. For full-duplex radios that can transmit and receive simultaneously, the interference coefficient among corresponding links can be assigned to zero.

\subsection{Optimization Problem Formulation}

Considering the conflicting objectives of throughput enhancement and energy saving, we will take a multi-objective optimization approach, which is to keep one objective and transform the other one to constraint \cite{liu2014energy,oh2013dynamic,luo2015downlink}. Particularly, the energy efficiency is optimized by minimizing the total energy consumption in the network while satisfying flow demands of multiple commodities.

Suppose the demand of commodity flow $\lambda$ is $q^{(\lambda)}$. To specify the source and destination of each flow, define demand vector $\bold{q}^{(\lambda)}=(q_1^{(\lambda)},\dots ,q_{|\mathcal{N}|}^{(\lambda)})'$ as
\begin{equation*}
  q_i^{(\lambda)}\triangleq
  \begin{cases}
   q^{(\lambda)}, & \text{if $i$ is the source node of flow $\lambda$}\\
   -q^{(\lambda)}, & \text{if $i$ is the destination node of flow $\lambda$}\\
  0,  & \text{otherwise}
  \end{cases}
\end{equation*}

Denote $f_l^{(\lambda)}$ as flow rate for commodity $\lambda$ on link $l$, then $\bold{f}^{(\lambda)}=(f_1^{(\lambda)},\dots ,f_{|\mathcal{L}|}^{(\lambda)})'$ is the flow vector of commodity $\lambda$ on all the links. Since flow rate is defined for each link, it can be related to nodes with an $|\mathcal{N}|$ by $|\mathcal{L}|$ node-flow incident matrix $\bold{H}$, whose entries are defined as
\begin{equation*}
  h_{ij}\triangleq
  \begin{cases}
   1, & \text{if link $j$ carries outgoing flow from node $i$}\\
   -1, & \text{if link $j$ carries incoming flow to node $i$}\\
  0,  & \text{otherwise.}
  \end{cases}
\end{equation*}

Based on the above definition, the flow balance constraints for each commodity can be expressed as
\begin{equation}\label{eq:flow_balance}
\bold{Hf}^{(\lambda)}=\bold{q}^{(\lambda)}, \quad \lambda=1,\cdots ,\Lambda
\end{equation}

Generally, a link may use different transmit power at different time such that the mutual interference among links can be dynamically coordinated and the transmission rate can be adjusted. At a time instance, the transmit power levels of all the tuple-links form a \emph{transmission pattern} (TP). A TP implies the transmission state of all the links in the network, including which radios and channels are being used as well as the corresponding transmit power and link capacity. Recall that the scheduling problem we defined is to decide when and how long the links should transmit such that the flow demands can be satisfied with minimum energy consumption. Therefore, with the concept of TP introduced, the problem of joint scheduling and power control becomes to select TPs and decide transmission time for them.

Since the sets of links and transmit power levels are finite, the total number of possible TPs is finite. In each TP, if a link is assigned a non-zero transmit power level, the tuple-link is considered to be \emph{active}. Let $\mathcal{A}$ be the set of all TPs in the network. Denote the portion of transmission time assigned to pattern $\alpha$ as $t_{\alpha}$. Let the transmit power and the capacity of link $l$ achieved in pattern $\alpha$ be $p_{l,\alpha}$ and $r_{l,\alpha}$, respectively. Since each TP defines the transmit power levels of all links, $r_{l,\alpha}$ can be expressed as a function of $p_{l,\alpha}$, which is
\begin{equation}\label{eq:rate}
r_{l,\alpha}=B_l \log_2(1+\frac{g_{l}p_{l,\alpha}}{\sum_{l' \in \mathcal{L}\setminus l}g_{l'l}p_{l',\alpha}+\sigma^2})
\end{equation}
Accordingly, the total traffic rate (including all commodities) on link $l$ is bounded as
\begin{equation}\label{eq:capacity_region}
f_l=\sum_{\lambda\in \Lambda}f_l^{(\lambda)}\le \sum_{\alpha\in\mathcal{A}}r_{l,\alpha}t_{\alpha}, \quad \forall l\in \mathcal{L}
\end{equation}
\begin{equation}\label{eq:time_constraint}
\sum_{\alpha\in\mathcal{A}}t_{\alpha}=1
\end{equation}

Thus, the energy-efficient resource allocation problem can be formulated as a TP based scheduling problem to minimize power consumption while satisfying flow demand, i.e.,

\indent {\bf Problem 1} (Original optimization problem):
\begin{align}
\min_{\{f_l^{(\lambda)},t_{\alpha}\}} & \quad E=\sum_{\alpha\in\mathcal{A}}\sum_{l\in\mathcal{L}}p_{l,\alpha} t_{\alpha} \label{eq:objl}\\
\textit{s.t.} & \quad\text{constraints (\ref{eq:flow_balance}),(\ref{eq:rate}),(\ref{eq:capacity_region}),(\ref{eq:time_constraint})} \nonumber\\
&\quad f_l^{(\lambda)}\geq 0, \quad \forall l\in\mathcal{L}, \quad \lambda=1,\cdots ,\Lambda \label{f_nonnegative}\\
&\quad x_{\alpha}\geq 0, \quad \forall \alpha\in\mathcal{A}
\end{align}
The optimization variables are flow variables $f_l^{(\lambda)}$, as well as transmission time portion $t_{\alpha}$ assigned to TPs. The objective function in (\ref{eq:objl}) stands for the total power consumption which is the summation of power consumption over all the links in all TPs.

\begin{lemma}\label{lemma:eq}
In the optimal solution of Problem 1, constraint (\ref{eq:capacity_region}) will reach equality.
\end{lemma}

\begin{IEEEproof}
The lemma can be proved by contradiction. Suppose with the optimal solution, constraint (\ref{eq:capacity_region}) does not reach equality, i.e., there exists a link $l$ such that $f_l<\sum_{\alpha\in\mathcal{A}}r_{l,\alpha}t_{\alpha}$. We call such a link over-scheduled, which indicates some pattern is providing more than necessary capacity to link $l$. Since $f_l$ is non-negative, there must exist a pattern $\alpha_1$ with $r_{l, \alpha_1}t_{\alpha_1}>0$ ($p_{l,\alpha_1}>0$).

Then we look for a pattern $\alpha_2$ that has smaller capacity on $l$ but no lower rate on the other links. In other words, pattern $\alpha_2$ should satisfy $0\le r_{l, \alpha_2}<r_{l, \alpha_1}$ and $r_{l', \alpha_2}\ge r_{l', \alpha_1},\forall l'\neq l$. It can be seen that any pattern with lower power level on $l$ and same levels on the other links applies. Since $p_{l,\alpha_1}>0$, we can always find such patterns. Notice that $\alpha_2$ has less power consumption than $\alpha_1$.

The equality on link $l$ can be achieved by removing the over-scheduled capacity on link $l$, which can be done by moving part of the traffic load from  $\alpha_1$ to $\alpha_2$. In other words, the equality can be achieved by designing a new schedule that moves a portion of $t_{\alpha_1}$ to $t_{\alpha_2}$. Such a new schedule $\{t'_{\alpha}\}$ can be obtained by solving:
\begin{align}
&t'_{\alpha_1}+t'_{\alpha_2}=t_{\alpha_1}+t_{\alpha_2} \\
&t'_{\alpha}=t_{\alpha}, \forall \alpha \in \mathcal{A}\setminus \{\alpha_1,\alpha_2\} \\
&f_l=\sum_{\alpha\in\mathcal{A}}r_{l,\alpha}t'_{\alpha}.
\end{align}

Under the new schedule, the capacity of other links will not be reduced while the inequality on link $l$ will become equality, which means Constraint (\ref{eq:capacity_region}) still holds and the new schedule is a feasible solution to Problem 1.

Since part of the transmission time of pattern $\alpha_1$ is re-scheduled to pattern $\alpha_2$ while the latter has less power consumption, the new scheduling solution will consume less power compared to the original one, which means the original solution is not optimal and contradicts the assumption. This completes the proof.
\end{IEEEproof}

The physical meaning of Lemma \ref{lemma:eq} is that whenever there is an over-scheduled link $l$, we can always adjust the scheduling to remove the redundant capacity by averaging out the traffic load from a current pattern to others with smaller capacity on $l$.

Then, with all the constraints in equality, we can rewrite the optimization problem into standard matrix form as follows:

\indent {\bf Problem 1} (Original problem in matrix form):
\begin{align*}
\min_{\bold{x}} & \quad \bold{c}' \bold{x} \\
\textit{s.t.} & \quad \bold{A} \bold{x} = \bold{b} \\
              &\quad \bold{x} \geq \bold{0}
\end{align*}
with $\bold{x}=(\bold{f}^{(1)'},\dots ,\bold{f}^{(\Lambda)'},t_1,\dots ,t_{|\mathcal{A}|})'$ and
\begin{align*}
\bold{A} & =\left (
 \begin{array}{ccc|c}
  \bold{H} &  &  &  \\
   & \ddots &  &  \\
   &  & \bold{H} &   \\
   -\bold{I}_{|\mathcal{L}|}& \cdots & -\bold{I}_{|\mathcal{L}|} & \bold{R}  \\
   &  &  & \bold{1}_{1\times |\mathcal{A}|}
 \end{array} \right )\\
\bold{b} & = (\bold{q}^{(1)'},\dots ,\bold{q}^{(\Lambda)'},\bold{0}_{1\times |\mathcal{L}|},1)' \\
\bold{c} & = (\bold{0}_{1\times (|\mathcal{L}|\Lambda)},\sum_{l\in\mathcal{L}}p_{l,1},\cdots, \sum_{l\in\mathcal{L}}p_{l,|\mathcal{A}|})'
\end{align*}
where $\bold{R}$ is the $|\mathcal{L}|\times |\mathcal{A}|$ link capacity matrix with entries $r_{l,\alpha}$. Notice that the non-zero entries in $\bold{c}$ correspond to the energy consumption of TPs.

It can be seen that Problem 1 is an LP problem; however, since the transmission patterns can be significantly many, searching the optimal schedule of the patterns across such a large solution space is difficult, which motivated us to develop a decomposition method to efficiently find the solution.

\section{Decomposition Framework}
The complexity of Problem 1 is mainly determined by the size of the TP set $\mathcal{A}$. For example, if all nodes have the same number of radios $|\mathcal{R}_v|$, the size of $\mathcal{A}$ is roughly $|\mathcal{A}|=|\mathcal{P}|^{|\mathcal{L}|}\approx |\mathcal{P}|^{(|\mathcal{N}|^2\cdot|\mathcal{R}_v|^2\cdot|\mathcal{C}|)}$, which will be significantly large.

Intuitively, not all the TPs will contribute to flow delivery and there is no need to allocate transmission time to TPs with little contribution. Our experiments in tuple-link scheduling \cite{cheng2014systematic,chengsystematic} also indicate that only a subset of $\mathcal{A}$ will be scheduled. In other words, instead of considering all the patterns in $\mathcal{A}$, we only need to find the critical ones and focus on the scheduling of these patterns. To this end, we develop a decomposition technique based on delayed column generation (DCG)\cite{cheng2014systematic} to iteratively find such a subset of critical TPs.

\subsection{DCG-Based Decomposition}\label{sec:decomposition}

According to the matrix form of Problem 1, the size of the left half of constraint matrix $\bold{A}$ is determined by the network topology, while each column in the right half corresponds to a TP. Therefore the number of scheduled TPs is equal to the number of columns in the right half of $\bold{A}$. Starting from an initial feasible solution obtained from a small subset of $\mathcal{A}$, the DCG method iteratively searches for new columns (or equivalently TPs) that are promising in improving the objective.

Let $\mathcal{A}^{(k)}$ denote the subset of TPs already found at the beginning of Step $k$. In Step $k$, the optimal solution with given $\mathcal{A}^{(k)}$ can be obtained by solving the following master problem:

\indent {\bf Problem 2} (Master Problem):
\begin{align}
\min_{\{f_l^{(\lambda)},t_{\alpha}\}} & \quad E^{(k)}=\sum_{\alpha\in \mathcal{A}^{(k)}}\left(\sum_{l\in\mathcal{L}}p_{l,\alpha}\right) t_{\alpha}, \\
{\it s.t.} & \quad  f_l=\sum_{\lambda\in \Lambda}f_l^{(\lambda)}= \sum_{\alpha\in\mathcal{A}^{(k)}}r_{l,\alpha}t_{\alpha} \quad \forall l\in \mathcal{L}\label{eq:constraint2}\\
& \quad\sum_{\alpha\in\mathcal{A}^{(k)}}t_{\alpha}=1 \\
&\quad x_{\alpha}\geq 0, \quad \forall \alpha\in\mathcal{A}^{(k)}\\
&\quad \text{constraints (\ref{eq:flow_balance}),(\ref{f_nonnegative})} \nonumber
\end{align}

Or in matrix form,

\indent {\bf Problem 2} (Master problem in matrix form):
\begin{align*}
\min_{\bold{x}^{(k)}} & \quad \bold{c}^{(k)'} \bold{x}^{(k)} \\
\textit{s.t.} & \quad \bold{A}^{(k)} \bold{x}^{(k)} = \bold{b} \\
              &\quad \bold{x}^{(k)} \geq \bold{0}
\end{align*}
where $\bold{c}  = (\bold{0}_{1\times (|\mathcal{L}|\Lambda)},\sum_{l\in\mathcal{L}}p_{l,1},\cdots, \sum_{l\in\mathcal{L}}p_{l,|\mathcal{A}^{(k)}|})$ and
$\bold{x}=(\bold{f}^{(1)'},\dots ,\bold{f}^{(\Lambda)'},t_1,\dots ,t_{|\mathcal{A}^{(k)}|})'$.
In the master problem, $\bold{A}^{(k)}$ has the same number of rows as $\bold{A}$ but much fewer columns than $\bold{A}$.

The above master problem can be easily solved if the subset $\mathcal{A}^{(k)}$ is of moderate size. The solution of the master problem provides the scheduling time $t_{\alpha}^{(k)}$ for each pattern $\alpha$ in $\mathcal{A}^{(k)}$ along with the dual variable vector $\bold{w}^{(k)}$ associated with the constraints (where $\bold{w}^{(k)}$ is obtained by solving the dual problem of Problem 2). The next problem is to search for a new column $\bold{A}_i$ to be added into $\mathcal{A}^{(k)}$ to generate $\mathcal{A}^{(k+1)}$, which can improve the objective of the optimization problem. In DCG algorithm, such an improvement is evaluated by the reduced cost $c_i-\bold{w}^{(k)'}\bold{A}_i$ where $i$ denotes the index of the new column \cite{bertsimas1997introduction}. If a column is associated with negative reduced cost, then adding this column will improve the objective value.

Since the added columns only correspond to the right half of constraint matrix, it can be observed that adding a column is equivalent to adding a new pattern $\alpha$, whose improvement\footnote{We use the additive inverse of reduced cost as a measurement of the performance improvement to keep consistency with the later definition of utility.} can be evaluated as
\begin{align}
&\bold{w}^{(k)'}\bold{A}_\alpha-c_\alpha \\
    =\;&0+\sum_{l\in\mathcal{L}}w_{l}^{(k)}r_{l,\alpha} +w_0^{(k)}\times 1 - \sum_{l\in\mathcal{L}}p_{l,\alpha} \nonumber\\
    =\;& \sum_{l\in\mathcal{L}}\left(w_{l}^{(k)}r_{l,\alpha} - p_{l,\alpha}\right)+w_0^{(k)} \label{eq:improvement}
\end{align}
where $w_{l}^{(k)}$ is the dual variable corresponding to the $l$'th row of matrix $\bold{R}^{(k)}$ in $\bold{A}^{(k)}$ (entry $r_{l,\alpha}$) and $w_0^{(k)}$ is the dual variable associated with the last row of $\bold{A}$.

Define the term $w_{l}^{(k)}r_{l,\alpha} - p_{l,\alpha}$ as the \emph{utility} of link $l$ in pattern $\alpha$, and the utility sum of all the links as \emph{system utility} $U^{(k)}$.
The utility of each link consists of the contribution to flow traffic and the power cost, where the flow contribution of a link is further determined by both the link capacity $r_{l,\alpha}$ and the dual variable $w_{l}^{(k)}$.
The expression of utility function $w_{l}^{(k)}r_{l,\alpha} - p_{l,\alpha}$ indicates that it should have the same unit as $p_{l,\alpha}$, which is power, while $w_{l}^{(k)}$ acts as a price factor to convert throughput into welfare.

A new TP will be added to $\mathcal{A}^{(k)}$ if it maximizes the improvement in (\ref{eq:improvement}). Since $w_0^{(k)}$ is a constant independent of the TP to be added in Step $k$ for a given $\mathcal{A}^{(k)}$, it can be ignored during pattern selection. Then selecting a new TP is equivalent to solving the following problem:

\indent {\bf Problem 3} (Sub-Problem):
\begin{align}
\max_{\alpha\in\mathcal{A}\backslash\mathcal{A}^{(k)}} & \quad U^{(k)}_{\alpha} = \sum_{l\in\mathcal{L}}\left(w_{l}^{(k)}r_{l,\alpha} - p_{l,\alpha}\right) \label{eq:utility}
\end{align}

Since energy efficiency is a compound of both the benefit in flow contribution and cost in power consumption, the expression in (\ref{eq:utility}) naturally provides an evaluation function of a TP with these considerations. Therefore, the sub-problem can be interpreted as to search for the most energy-efficient TP, which is evaluated by the corresponding system utility. As mentioned previously, the system utility shares the same unit as that of power, which indicates that the objectives in sub-problem and original problem are consistent in unit.

The new TP, if found, is then added to $\mathcal{A}^{(k)}$ to form $\mathcal{A}^{(k+1)}$. The master problem is then updated and solved to provide a new set of solutions.
The process is repeated until no improvement can be made (or no column can be added), i.e., $\bold{w}^{(k)'}\bold{A}_i-c_i\le 0, \forall i$. Then, the standard DCG theory shows that the current solution will be the optimal solution of Problem 1 \cite{bertsimas1997introduction}.

The physical meaning behind the DCG decomposition can be explained as follows. We search for energy efficient TPs to perform scheduling, where the energy efficiency of TPs depends on the information obtained from current solution. Each time solving the master problem will provide an updated evaluation on all the links regarding their capabilities in satisfying traffic demand based on their performance in existing TPs, and such an evaluation is conveyed through dual variables $w_{l}^{(k)}$. Then according to this evaluation, the most energy efficient TP that can maximize the system utility is searched and fed back to the master problem. With this new information, all the links will be re-evaluated through solving the updated master problem. Repeating these steps will provide more and more accurate evaluations on the energy efficiency of TPs and therefore approach the optimal solution.

According to the definition of TP, finding a TP is equivalent to finding a power allocation over all the links. In this sense, the proposed framework can be interpreted as decomposing the original problem into scheduling phase (master problem) and power allocation phase (sub-problem). Optimality remains intact during the decomposition process by iteratively solving the two phases. Thus, with the multi-dimensional modeling, TP based scheduling and DCG based decomposition, the joint optimal solution over all dimensions of network resources can be obtained.

\subsection{Initial Solution}

It is usually difficult to find an initial subset of $\mathcal{A}$ that can yield feasible solution. However, even if the initial solution is infeasible, we can still find new columns based on the dual variables, and the newly added columns can potentially drive the iteration to yield a feasible solution. Thus, even starting from an infeasible solution, the feasibility will be restored in several rounds, providing that the original Problem 1 is feasible.

Based on the above observation and analysis, the initial subset can be constructed with randomly selected TPs. However, if a link is not included in the initial subset, then the link will probably never be evaluated or involved into the problem. Taken this issue into consideration, we need to cover every link in the initial subset. In addition, the constraint matrix should have full row rank \cite{bertsimas1997introduction}. Based on these, the initial subset can be set by choosing $|\mathcal{L}|$ TPs where each TP has exactly one unique link activated. In this way, we can get a diagonal matrix $\bold{R}$ and matrix $\bold{A}$ will have full row rank.

\subsection{Greedy Algorithm for Solving the Sub-Problem}

The sub-problem is to find a TP with maximum utility, which is done by searching over all the unused patterns. Again, the large searching space leads to impractical computational complexity.
As previously mentioned, a TP is defined as a power allocation on all the links, therefore it is equivalent to finding an optimal power allocation on links to maximize the system utility (the superscript indicating number of rounds is omitted in this sub-section since the sub-problem is solved within one round):
\begin{equation}
\max \limits _{\{p_l\}} \quad U=\sum_{l\in\mathcal{L}}u_l=\sum_{l\in\mathcal{L}}w_lr_l-p_l
\end{equation}

Power allocation on links with the objective of maximizing system utility is a challenging problem since the utilities of links are mutually dependent. Even if the utility of each link is fixed, the problem is still NP hard (which can be viewed as a maximum weighted independent set problem under physical interference model as in \cite{wan2011wireless}).
To obtain a practically feasible solution, we develop a greedy algorithm to find the optimal power allocation.

The greedy power allocation is done by starting with all-zero power allocation and gradually activating (assigning positive power levels to) links until the system utility no longer increases. Whether an inactive link can be activated depends on its contribution to the system utility. Among all the inactive links, the one with the largest contribution will be activated.

The details of the greedy algorithm are shown in Algorithm \ref{alg:greedy}. Denote the set of active links and inactive links as $\mathcal{S}_a$ and $\mathcal{S}_i$, respectively. At each step, the algorithm evaluates all the inactive links and selects one into the active set.
According to Eq. (\ref{eq:SINR}) and the definition of utility, the utility of each link can be written as a function of its transmit power $p_l$ and active link set $\mathcal{S}$,
\begin{align}
u_l&=u_l(p_l,\mathcal{S})\nonumber\\
&=w_lr_l-p_l \nonumber\\
&=w_l B_l \log_2 (1+\frac{g_{l}p_{l}}{\sum\limits_{l' \in
\mathcal{S}\setminus \{l\}}g_{l'l}p_{l'}+\sigma^2})-p_l \label{eq:u_l}
\end{align}
Each inactive link first calculates a myopic optimal power level $\hat{p}_l$ that maximizes its own utility assuming that the power levels of all the other links keep unchanged. $\hat{p}_l$ can be obtained by calculating the utilities at all possible power levels and choosing the one that gives maximum utility. Then it calculates the change of system utility $\Delta U_l$ if it is activated by using power $\hat{p}_l$,
\begin{equation}\label{eq:Delta_U}
\Delta U_l= u_l(\hat{p}_l,\mathcal{S}_a)+\sum\limits_{l'\in \mathcal{S}_a}u_{l'}(p_{l'},\mathcal{S}_a \cup \{l\})-\sum\limits_{l'\in \mathcal{S}_a}u_{l'}(p_{l'},\mathcal{S}_a)
\end{equation}
 i.e., $\Delta U_l$ can be viewed as the contribution of $l$ if activated. For the link with the largest contribution, if its contribution is larger than a pre-defined non-negative constant $\epsilon$\footnote{Normally $\epsilon$ can be set to 0. However, the value of $\epsilon$ can be positive if we want to terminate the algorithm earlier when the contribution of adding a new link is very small.}, it means activating this link can increase the system utility and this link will be activated at the calculated power level $\hat{p}_l$. Otherwise, the system utility cannot be increased and the algorithm stops. At the end of the algorithm, it outputs the power allocation to all links.

\begin{algorithm}[htbp]

 \caption{Greedy Algorithm for Problem 3}\label{alg:greedy}

\textbf{Input:} dual variables $\{w_l\}_{l=1,\cdots ,|\mathcal{L}|}$\;

\textbf{Initialization:} $p_l=0, \forall l\in \mathcal{L}$; $\mathcal{S}_a=\emptyset$; $\mathcal{S}_i=\mathcal{L}$\;

 \While{$\mathcal{S}_i\neq \emptyset$}{

 \For{$l\in \mathcal{S}_i$}{
 $\hat{p}_l= \arg\max u_l(p_l,\mathcal{S}_a)$\;
 Calculate $\Delta U_l$ according to Eq. (\ref{eq:Delta_U})\;
 }%end of for
 $l^*=\arg\max\limits_{l\in \mathcal{S}_i}\Delta U_l$ (If the solution is not unique, randomly select one link with $\max\limits_{l\in \mathcal{S}_i}\Delta U_l$)\;
 \eIf{$\Delta U_{l^*}>\epsilon$}{
 move $l^*$ from $\mathcal{S}_i$ to $\mathcal{S}_a$; $p_{l^*}=\hat{p}_{l^*}$\;
 }{
 Algorithm stops\;
 }
 }

\textbf{Output:} power allocation $\{p_l\}_{l=1,\cdots ,|\mathcal{L}|}$.

\end{algorithm}

\subsection{Complexity Analysis}

Each utility computation (as in Eq. (\ref{eq:u_l})) incurs a complexity in the order of $|\mathcal{S}_a|$. Within each iteration of the greedy algorithm, each inactive link in $\mathcal{S}_i$ will perform $|\mathcal{P}|$ utility computations to find the optimal power level and at most $2|\mathcal{S}_a|$ utility computations (as in Eq. (\ref{eq:Delta_U})) to calculate the effect on other links. As a result, each iteration requires a total number of $(|\mathcal{P}|+2|\mathcal{S}_a|)|\mathcal{S}_i||\mathcal{S}_a|$ computations to evaluate the contributions of all the inactive links, plus $|\mathcal{L}|$ computations to perform sorting. The iteration will be repeated by $|\mathcal{S}_a|$ times, leading to a total complexity of $[(|\mathcal{P}|+2|\mathcal{S}_a|)|\mathcal{S}_i||\mathcal{S}_a|+|\mathcal{L}|]|\mathcal{S}_a|$.

In practice, there can be at most $|\mathcal{R}|/2$ links actived simultaneously due to radio conflict. Therefore in the result of the algorithm we will have $|\mathcal{S}_a|\le |\mathcal{R}|/2$. In addition, $|\mathcal{S}_i|\le |\mathcal{L}|$. Based on this, the computation complexity is in the order of $(|\mathcal{R}|+|\mathcal{P}|)|\mathcal{R}|^2|\mathcal{L}|$. Further, $|\mathcal{R}|$ is usually less than $|\mathcal{L}|$. Therefore the greedy algorithm's complexity will be in the order of $|\mathcal{P}||\mathcal{L}|^3$.

Since the number of TPs is $|\mathcal{P}|^{|\mathcal{L}|}$ and the complexity of calculating system utility of each TP is $|\mathcal{L}|^2$, the complexity of brute force searching over the entire space to find maximum utility TP is $|\mathcal{L}|^2|\mathcal{P}|^{|\mathcal{L}|}$, which is significantly higher than that by Algorithm \ref{alg:greedy}.

\subsection{Algorithm Design}

With the decomposition framework and the greedy algorithm, we can now design the decomposition algorithm for solving the original problem, as shown in Algorithm \ref{alg:main}.

\begin{algorithm}[htbp]

 \caption{Decomposition Algorithm for Problem 1}\label{alg:main}
    Initial transmission pattern set $\mathcal{A}^{(0)}$\;

 \While{$E^{(k)}<E^{(k-1)}$}{

 //Master stage:

 Update master problem (Problem 2) with current TPs $\mathcal{A}^{(k)}$\;
 Solve master problem to obtain energy $E^{(k)}$ and dual variables $\bold{w}^{(k)}$\;

 //Sub-problem stage:

 Search for a new TP $\alpha$ by solving the sub-problem (Problem 3) using Algorithm \ref{alg:greedy} \;

 \eIf{$\bold{w}^{(k)'}\bold{A}_{\alpha}-c_{\alpha}>0$}{
 Add the new TP to $\mathcal{A}^{(k)}$ and obtain $\mathcal{A}^{(k+1)}$\;
 $k \leftarrow k+1$\;
 Go to master stage\;
 }{
 break\;
 }

}

\end{algorithm}

\section{Performance Analysis}

It is known that, theoretically, the DCG-based iterative algorithm will converge to an optimal solution of the original problem, providing that the sub-problem is optimally solved in every step \cite{bertsimas1997introduction}. Therefore, in our case, the optimality of the obtained solution is determined by that of the sub-problem. Below, we will first show how the performance of the greedy algorithm affects that of the original problem (Problem 1).

\subsection{Performance of the Original Problem Solution}

Denote the corresponding objectives achievably optimal solutions of the original problem and the sub-problem as $E^*$ and $U^*$, respectively. When Algorithm \ref{alg:main} stops, let the solution of the sub-problem solved by the greedy algorithm be $\hat{U}$, and the corresponding solution to Algorithm \ref{alg:main} be $\hat{E}$. Then we have the following relationship:

\begin{theorem} \label{theorem:gap}
The performance gap of Algorithm \ref{alg:main} in solving the original problem is upper bounded by that of Algorithm \ref{alg:greedy} in solving the sub-problem, i.e.,
\begin{equation}
\Delta_E=\hat{E}-E^* \le U^*-\hat{U}=\Delta_U
\end{equation}
\end{theorem}

For the original problem, suppose the dual vector associated with $\hat{E}$ is $\hat{\bold{w}}$, whose last entry is $\hat{w}_0$. Before proving Theorem \ref{theorem:gap}, we first present the following result.
\begin{lemma} \label{lemma:feasibility}
If $\hat{\bold{w}}$'s last entry ($\hat{w}_0$) is replaced by $\hat{w}_0-\Delta_U$, the resulting vector, denoted as $\tilde{\bold{w}}$, will still be a feasible solution to Problem 1's dual problem.
\end{lemma}

\begin{IEEEproof}
Denote the dual problem of Problem 1 as:

\indent {\bf Problem 1D} (dual of Problem 1):
\begin{align*}
\max_{\bold{w}} &\quad \bold{w}'\bold{b} \\
{\it s.t.} &\quad \bold{w}'\bold{A} \le \bold{c}'
\end{align*}

Denote $\bold{A}_i$ as a column of $\bold{A}$. For the columns in the left half of $\bold{A}$, replacing $\hat{w}_0$ with $\hat{w}_0-\Delta_U$ will not affect the value of $\hat{\bold{w}}'\bold{A}_i$ since the last row of left half of $\bold{A}$ only has zero entries. Therefore, for these columns, $\tilde{\bold{w}}'\bold{A}_i \le c_i$ still holds.

Recalling that each column of the right half of $\bold{A}$ is associated with a TP, we can write Eq. (\ref{eq:improvement}) as $\bold{w}'\bold{A}_\alpha-c_{\alpha}=U_{\alpha}+\hat{w}_0$, where $U_{\alpha}$ is the system utility achieved by pattern $\alpha$.

Since the decomposition algorithm stops at $\hat{U}$, we have
\begin{equation*}
\hat{U}+\hat{w}_0=\hat{\bold{w}}'\bold{A}_\alpha-c_\alpha \le 0
\end{equation*}
which indicates
\begin{equation*}
\hat{w}_0-\Delta_U \le -U^*
\end{equation*}

For $\tilde{\bold{w}}$ and every column in the right half of $\bold{A}$, we have
\begin{align*}
\tilde{\bold{w}}'\bold{A}_\alpha-c_{\alpha}&=U_{\alpha}+\hat{w}_0-\Delta_U \\
&\le U_{\alpha}-U^* \\
&\le 0
\end{align*}

Above all, $\tilde{\bold{w}}$ is a feasible solution to Problem 3.
\end{IEEEproof}

Then we continue to prove Theorem \ref{theorem:gap}.
\begin{IEEEproof}
Suppose $\bold{x^*}$ and $\bold{w^*}$ are the optimal solutions of the original problem (Problem 1) and its dual problem (Problem 1D), respectively. From Lemma \ref{lemma:feasibility}, $\tilde{\bold{w}}$ is a feasible solution to Problem 1D. Therefore
\begin{equation*}
\bold{w^*}'\bold{b} \ge \tilde{\bold{w}}'\bold{b}= \hat{\bold{w}}'\bold{b}-\Delta_U =\hat{E}-\Delta_U
\end{equation*}

According to weak duality, $\bold{c}\bold{x^*} \ge \bold{w^*}'\bold{b} $, which leads to
\begin{align*}
\hat{E}-E^*=\hat{E}-\bold{c}\bold{x^*} &\le\hat{E}-\bold{w^*}'\bold{b} \\
&\le \hat{E}-(\hat{E}-\Delta_U)\\
&=\Delta_U
\end{align*}
thus completing the proof of Theorem \ref{theorem:gap}.
\end{IEEEproof}

As discussed in Section \ref{sec:decomposition}, it can be observed that both performance gaps of the solutions of original problem and sub-problem are in the unit of power, which shows the consistency of unit in Theorem \ref{theorem:gap}. Moreover, Theorem \ref{theorem:gap} shows that the performance gap of the original optimization problem¡¯s solution is bounded by that of the sub-problem. Therefore, the performance of the decomposition algorithm can be evaluated through investigating the performance gap of Algorithm \ref{alg:greedy} in solving the sub-problem.

\subsection{Performance of the Sub-Problem Solution}\label{sec:bound}

The objective of the sub-problem is the system utility, whose maximum value is related to how many links can be activated simultaneously. Due to radio conflict, links sharing the same radio will not be activated at the same time, otherwise both of them will result in 0 utility. Considering that the tuple-link based multi-dimensional network model can be abstracted as a graph with radios being vertices and links being edges, the maximum number of concurrent links with positive utility can be characterized by the matching number of the graph associated with the network. Let $M^*$ denote the matching number of the network. We have the following statement.

\begin{lemma}\label{lemma:matching}
In the optimal solution of the sub-problem, there can be at most $M^*$ links with positive utility.
\end{lemma}
The proof of Lemma \ref{lemma:matching} follows directly the definition of matching number of graph.

Define $\tilde{u}_l=u_l(\hat{p}_l,\emptyset)$ as the \emph{utopian utility} of a link, which is its optimal utility when ignoring any mutual interference. Notice that utopian utility is also the utility of each link in the first round of the greedy algorithm (Algorithm \ref{alg:greedy}), and will not be smaller than the practically achieved utility when the corresponding link is scheduled. Without loss of generality, suppose $\{\tilde{u}_l\}$ is sorted in descending order, i.e., $\tilde{u}_1>\tilde{u}_2>\cdots$. Based on this, we can derive one performance bound of Algorithm \ref{alg:greedy} as

\begin{lemma}\label{lemma:constant}
\begin{equation}
\frac{\hat{U}}{U^*} \ge \frac{\tilde{u}_1}{\sum\limits_{l=1}^{M^*}\tilde{u}_l} \ge \frac{1}{M^*}
\end{equation}
\end{lemma}
\begin{IEEEproof}
According to Algorithm \ref{alg:greedy}, the system utility will be increased every time a new link is added, therefore the final system utility of the greedy algorithm $\hat{U}$ will not be smaller than that in the first round, i.e., $\hat{U}\ge \tilde{u}_1$. On the other hand, there can be at most $M^*$ links with positive utility according to Lemma \ref{lemma:matching}. Hence, $U^* \le \sum\limits_{l=1}^{M^*}\tilde{u}_l$. Together we have
 \begin{equation*}
\frac{\hat{U}}{U^*}
\ge \frac{\tilde{u}_1}{\sum\limits_{l=1}^{M^*}\tilde{u}_l}
\ge \frac{\tilde{u}_1}{M^*\tilde{u}_1}
= \frac{1}{M^*}
\end{equation*}
\end{IEEEproof}

Lemma \ref{lemma:constant} shows that the performance of the greedy algorithm is constant-bounded, and implies two ways of evaluating performance gap $\Delta_U$, which are shown in the following theorem.
\begin{theorem}\label{theorem:constant}
The performance gap of Algorithm \ref{alg:greedy} in solving sub-problem is upper bounded as
\begin{align}
\Delta_U &\le \sum\limits_{l=2}^{M^*}\tilde{u}_l \label{eq:ubound}\\
&\le (M^*-1)\hat{U} \label{eq:Rbound}
\end{align}
\end{theorem}

\subsection{Bound from Sub-Problem Relaxation}\label{sec:relax}

For a given $\hat{U}$, estimating the upper bound of $\Delta_U$ is equivalent to estimating the upper bound of $U^*$. An upper bound of U* can be obtained by solving a relaxed version of Problem 3 as follows.

The problem can be relaxed by ignoring some interference without changing the formulation. In other words, the relaxation of Problem 3 can be done by reducing the values of $g_{ll'}$'s. For example, one relaxation can be ignoring all the interference or radio conflict in the network (i.e., $g_{ll'}=0, \forall l,l' \in \mathcal{L}$) but limiting the total number of active links to $M^*$. In this case, the utility of each link is independent of other links' activities and the optimal system utility is $\sum\limits_{l=1}^{M^*} \tilde{u}_l$, which is an interpretation of the bound in Eq. (\ref{eq:ubound}). However, in this case, tuple-links associated with the same physical link usually have the same dual values ($w_l$), which means they tend to be activated simultaneously if mutual interference is ignored. As a result, in the solution of this relaxed optimization problem, many links share same radios, which is physically infeasible. Therefore, this relaxation may yield a loose bound.

Another relaxation is to ignore only the co-channel interference by setting $g_{ll'}=0$ to links not having radio conflict, while keeping the other $g_{ll'}$'s unchanged. In fact, this relaxation will result in a maximum weighted matching problem, whose optimal solution is still hard to find.

In order to formulate a proper relaxed problem to characterize an upper bound of $U^*$, we consider a point between the previously mentioned two examples. This relaxation will ignore co-channel interference and modify the radio constraint as follows: There could be at most $\min\{\mathcal{R}_u,\mathcal{R}_v\}$ tuple-links activated on any physical link between node $u$ and $v$,  while the total number of active tuple-links in the network is limited by $M^*$. Denote the relaxed problem as {\bf Problem 3R}. The solution of Problem 3R can be obtained by greedily picking $\{\tilde{u}_l\}$ as long as the above constraint is not violated, as summarized in Algorithm \ref{alg:sub}.

\begin{algorithm}[htbp]

 \caption{Solving Problem 3R}\label{alg:sub}
\textbf{Input:} $\{\tilde{u}_l\}$ (sorted in descending order)\;

\textbf{Initialization:}

Number of selected tuple-links on each physical link $n_{uv}=0, \forall u,v\in \mathcal{N}$\; Total Number of active tuple-links $n=0$\; Total utility $U=0$\; $l=1$\;

 \While{$n<M^*$ and $\tilde{u}_l>0$}{
 Find $l$'s corresponding physical link $uv$\;
 \If{$n_{uv}<\min\{\mathcal{R}_u,\mathcal{R}_v\}$}{
 Tuple-link $l$ is activated;
 $n=n+1$\;
 $n_{uv}=n_{uv}+1$\;
 $U=U+\tilde{u}_l$\;
 }
 $l=l+1$\;
 }%end of while

\textbf{Output:} $U$ as the solution of Problem 3R.

\end{algorithm}

\begin{lemma}
Algorithm \ref{alg:sub} yields the optimal solution of Problem 3R.
\end{lemma}
\begin{IEEEproof}
	Since the constraint only applies to each physical link locally and there is no co-channel interference, the behavior of each physical link has no influence on other physical links. On the other hand, it can be observed that the greedy selection of the links associated with one physical link is locally optimal. As a result, combining the local optimal solutions of independent physical links will yield the global optimal solution.
\end{IEEEproof}

Define the list of utopian utilities as utopian list ($\{\tilde{u}_l\}, l=1,2,\dots ,\mathcal{L}$). In Algorithm \ref{alg:sub}, at most $M^*$ utopian utilities are added to the final system utility. Notice that these added utilities may not be the first $M^*$ ones ($M^*$ largest ones) in the utopian list, since some might be excluded due to the constraint defined in Problem 3R. Then define the list of added utopian utilities associated with the selected links in Algorithm \ref{alg:sub} as reduced utopian list ($\{\tilde{u}^R_l\},l=1,2,\dots ,M^*$). From the previous analysis, it can be observed that each $\tilde{u}^R_l$ will be no larger than the corresponding $\tilde{u}_l$ at the same position in the list\footnote{Here the subscript $l$ indicates the location in the list, not a link. That is, $\tilde{u}^R_l$ and $\tilde{u}_l$ are at the same location in the corresponding lists, but not necessarily belong to the same link.}, that is
\begin{equation} \label{eq:reduced}
\tilde{u}^R_l \le \tilde{u}_l, \forall 1\le l \le M^*
\end{equation}
Denote the optimal solution of Problem 3R as $\tilde{U}$, then Theorem \ref{theorem:3R} directly follows:
\begin{theorem}\label{theorem:3R}
The performance gap of Algorithm \ref{alg:greedy} in solving sub-problem is upper bounded as
\begin{align}
\Delta_U &\le \tilde{U}-\hat{U} \le \sum\limits_{l=2}^{M^*}\tilde{u}^R_l
\end{align}
where $\tilde{u}^R_l$ are the utilities from reduced utopian list.
\end{theorem}

Compared with (\ref{eq:ubound}) in Theorem \ref{theorem:constant}, the right hand sides are the summations of the same number of entries, while each entry in that of Theorem \ref{theorem:3R} is no larger than that of Theorem \ref{theorem:constant} (as implied by (\ref{eq:reduced})), therefore the bound in Theorem \ref{theorem:3R} is usually tighter.

\section{Numerical Results}\label{sec:performance}

We consider a connected MR-MC network with 25 nodes randomly deployed in a 1000 $\times$ 1000 m$^2$ area. Each node is equipped with one or multiple radio interfaces, and there are multiple channels available for all radios. The transmit power of each radio can take values on a logarithmic scale from 0 to $p_{\text{max}}$. The parameter settings are listed in Table \ref{parameter}.

\begin{table}[!t]
\renewcommand{\arraystretch}{1.3}
\caption{Parameter Setting}
\label{parameter}
\centering
\begin{tabular}{c|c}
\hline
\hline
Parameter & Value (Default)\\
\hline
$p_{\text{max}}$ & 10mW \\
Maximum transmission range &  250m \\
Channel noise power &  -30dBm \\
Path loss factor &  2 \\
Flow demand &  35 or 70 Kbps per flow \\
Number of radios & 1-3 per node (2)  \\
Number of channels & 1-8 (4)  \\
Bandwidth & 1Mbps  \\
\hline
\hline
\end{tabular}
\end{table}

\subsection{Iteration and Optimality}

\begin{figure*}[htbp]
  \centering
        \subfigure[sample1.]{
            \includegraphics*[width=2.2in]{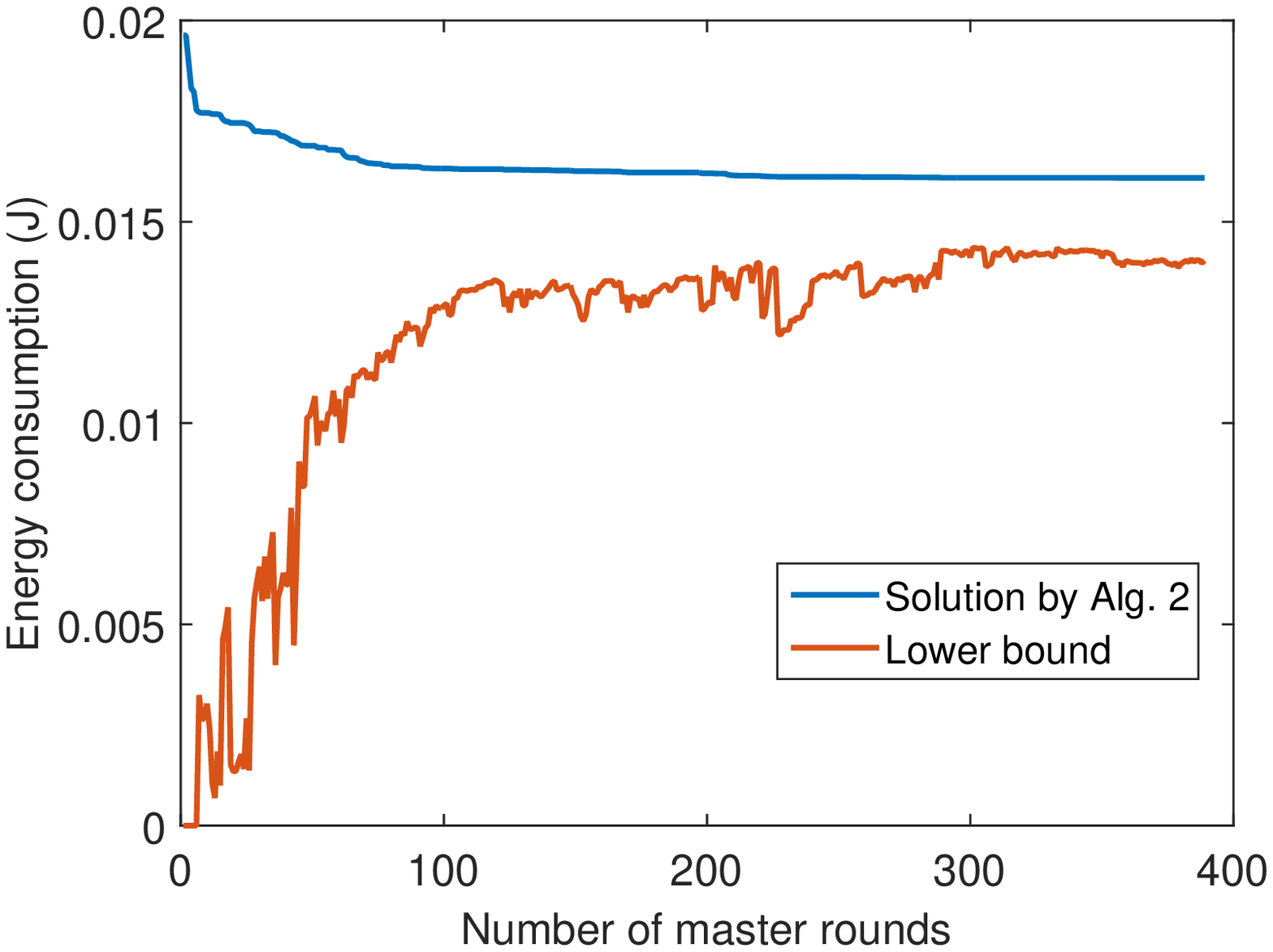}
        }
        \subfigure[sample2.]{
            \includegraphics*[width=2.2in]{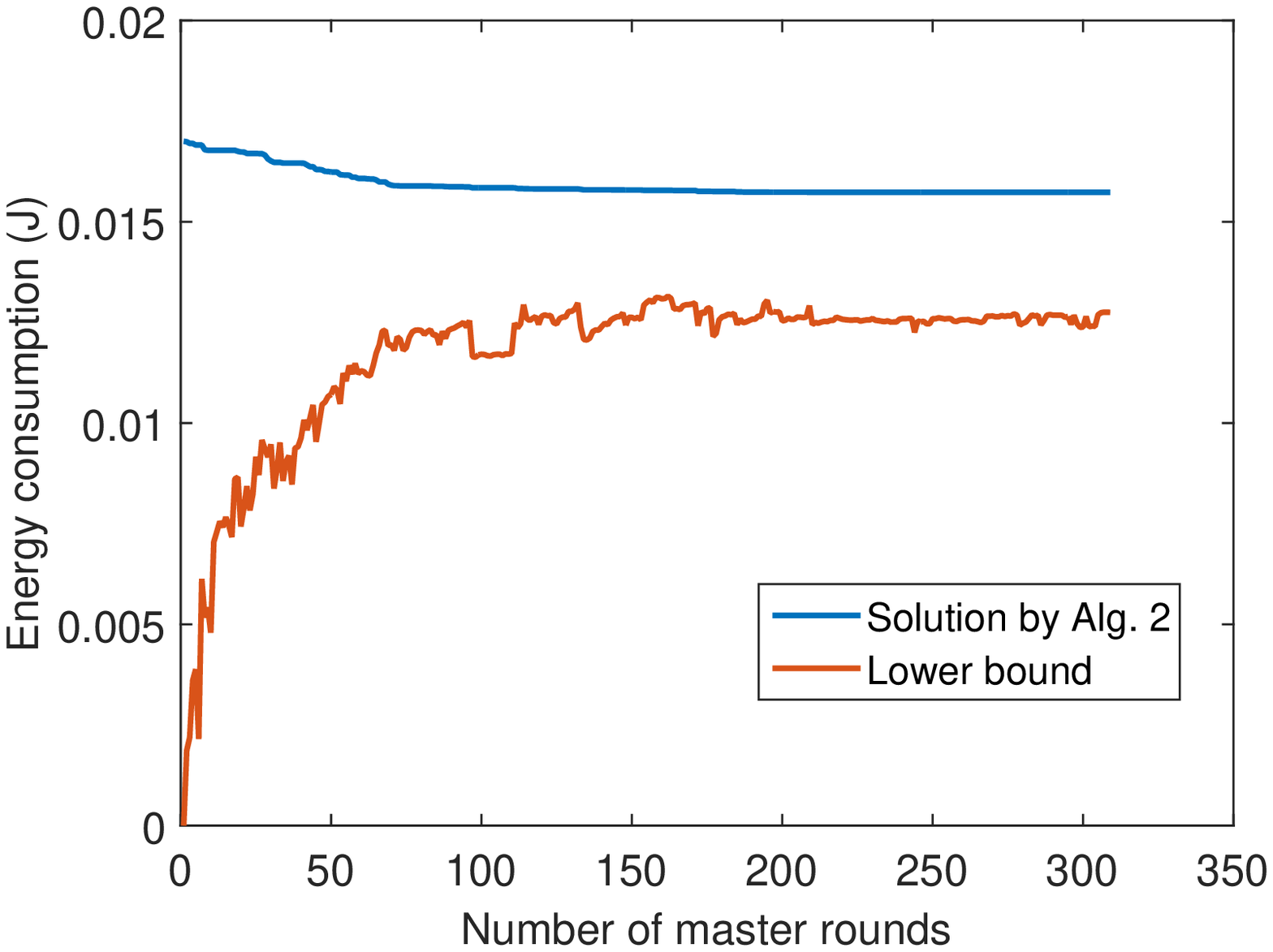}
        }
	    \subfigure[sample3.]{
            \includegraphics*[width=2.2in]{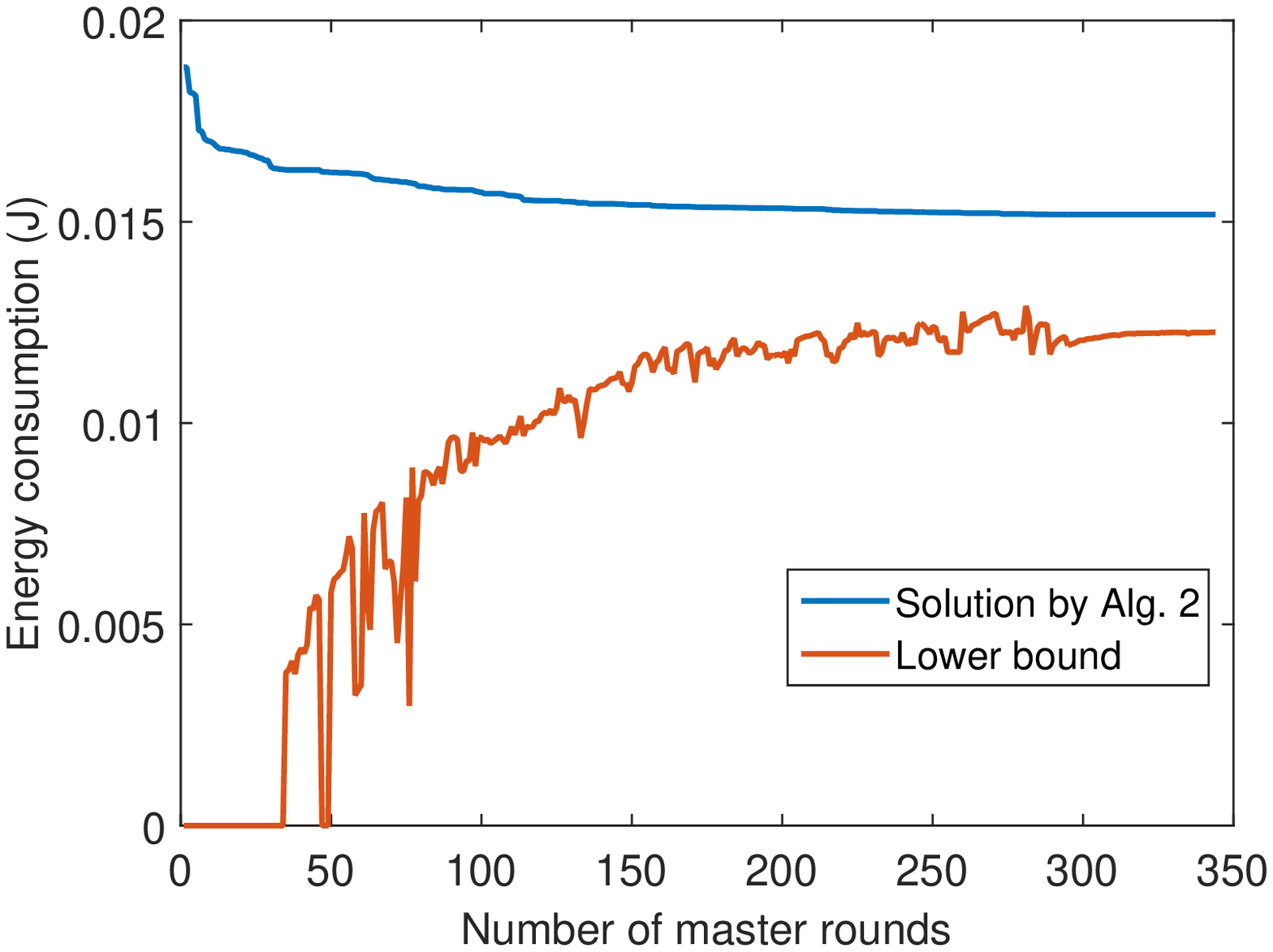}
        }
\caption{Energy consumption of the proposed algorithm and the lower bound of optimum.}\label{fig:bound}
\end{figure*}

We first tested the proposed decomposition framework and greedy algorithm on several sample topologies. In order to compare the result with the optimal solution, we solve the relaxed problem (Problem 3R) as described in Section \ref{sec:relax} and apply Theorem \ref{theorem:gap} to obtain the lower bound of the optimal solution\footnote{Theoretically the lower bound is obtained when the algorithm converges. However since we solved the relaxed problem in every round, the intermediate results are also presented as lower bound.}. The result of the objective value of Problem 1 (energy consumption) and the corresponding lower bound of optimum (with intermediate results) are shown in Fig. \ref{fig:bound}.

As shown in Fig. \ref{fig:bound}, when the result converges, there is a gap from the lower bound to the optimum. This gap naturally exists since the lower bound is calculated from the relaxed problem which ignores all co-channel interference and part of radio conflict. The optimal solution of the original problem will be worse (larger) than the lower bound, thus the actual distance between our solution and optimum will be smaller than the gap shown in the figures.

\subsection{Effect of Power Control}

One of our major contributions in this paper is the joint scheduling and power control by applying more realistic physical interference model in multi-dimensional resource allocation. We compare the performance of the joint optimization with that without power control. Energy efficiency of the network is used as the performance metric, which is defined as the ratio of sum traffic demands of all commodities and total energy consumption (the objective function of Problem 1).

\begin{figure}[ht]
  \centering
  \scalebox{0.45}
  {\includegraphics{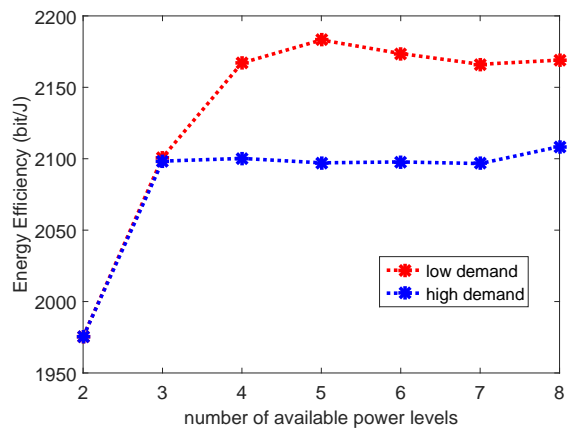}}
  \caption{Energy efficiency under different number of available power levels.}\label{fig:powerlevel}
\end{figure}

In order to demonstrate the effect on energy efficiency from joint optimization, we vary the number of available power levels and compare the achieved energy efficiency. Notice that when $|\mathcal{P}|=2$, transmit power can only be either zero or maximum transmit power, which can be viewed as the solution without power control. The energy efficiency corresponding to different number of available power levels $|\mathcal{P}|$ is shown in Fig. \ref{fig:powerlevel}. As can be seen from this figure, the proposed approach with power control ($|\mathcal{P}|>2$) always outperforms that without power control. This is because when without power control, whenever a link is scheduled for transmission, the maximum transmit power is used, which might be unnecessarily high and generates high interference to other links. Especially when the traffic demand is low, allowing links to transmit at low power levels can be beneficial in improving energy efficiency. A more fine-grained power level set can also increase the number of TPs and hence facilitates better resource allocation to reduce mutual interference. Therefore involving power control into joint resource allocation can improve the network energy efficiency.

On the other hand, as illustrated in Fig. \ref{fig:powerlevel}, further increasing the number of available power levels actually makes little difference to the network energy efficiency. For example, to meet a high traffic demand, links tend to use high transmit power in order to increase link capacity, leading to no use of low power levels. In view of the increased complexity of Algorithm \ref{alg:main} for more power levels, it is suggested to properly allocate the power level set according to the traffic demand level.

\begin{figure}[ht]
  \centering
  \scalebox{0.45}
  {\includegraphics{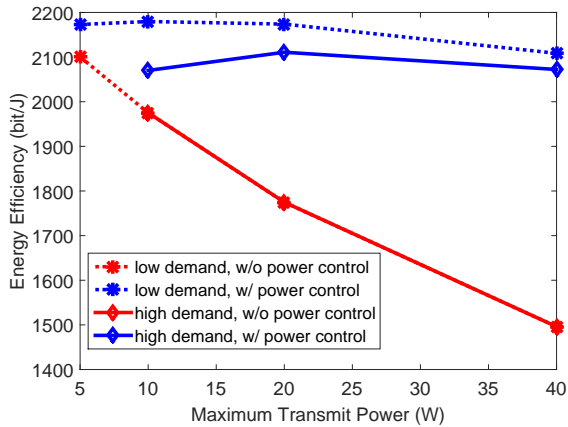}}
  \caption{Energy efficiency under different $p_{\text{max}}$.}\label{fig:pmax}
\end{figure}

We further compare the results under various maximum transmit power $p_{\text{max}}$ for the radios. As shown in Fig. \ref{fig:pmax}, without power control, we may observe a trend that the energy efficiency will decrease as $p_{\text{max}}$ increases. Similarly as previous discussion, when $p_{\text{max}}$ is increased, the fixed power case has to use higher power for transmission, leading to a degradation in energy efficiency. While in the case with power control, even if $p_{\text{max}}$ is increased, radios are still able to use low transmit power. As a result, the energy efficiency is almost unchanged when varying $p_{\text{max}}$. In practice, if a large $p_{\text{max}}$ has to be chosen in order to satisfy high traffic demand, then taking power control into resource allocation can help maintain the energy efficiency of the network, which shows another benefit of joint resource allocation.

\subsection{Sensitivity to Radio/Channel Resources}

We further evaluate the performance of the proposed joint resource allocation under different network configurations and investigate the effect of different types of network resources on energy efficiency.

\begin{figure*}[htbp]
  \centering
        \subfigure[1 radio per node.]{
            \includegraphics*[width=2.2in]{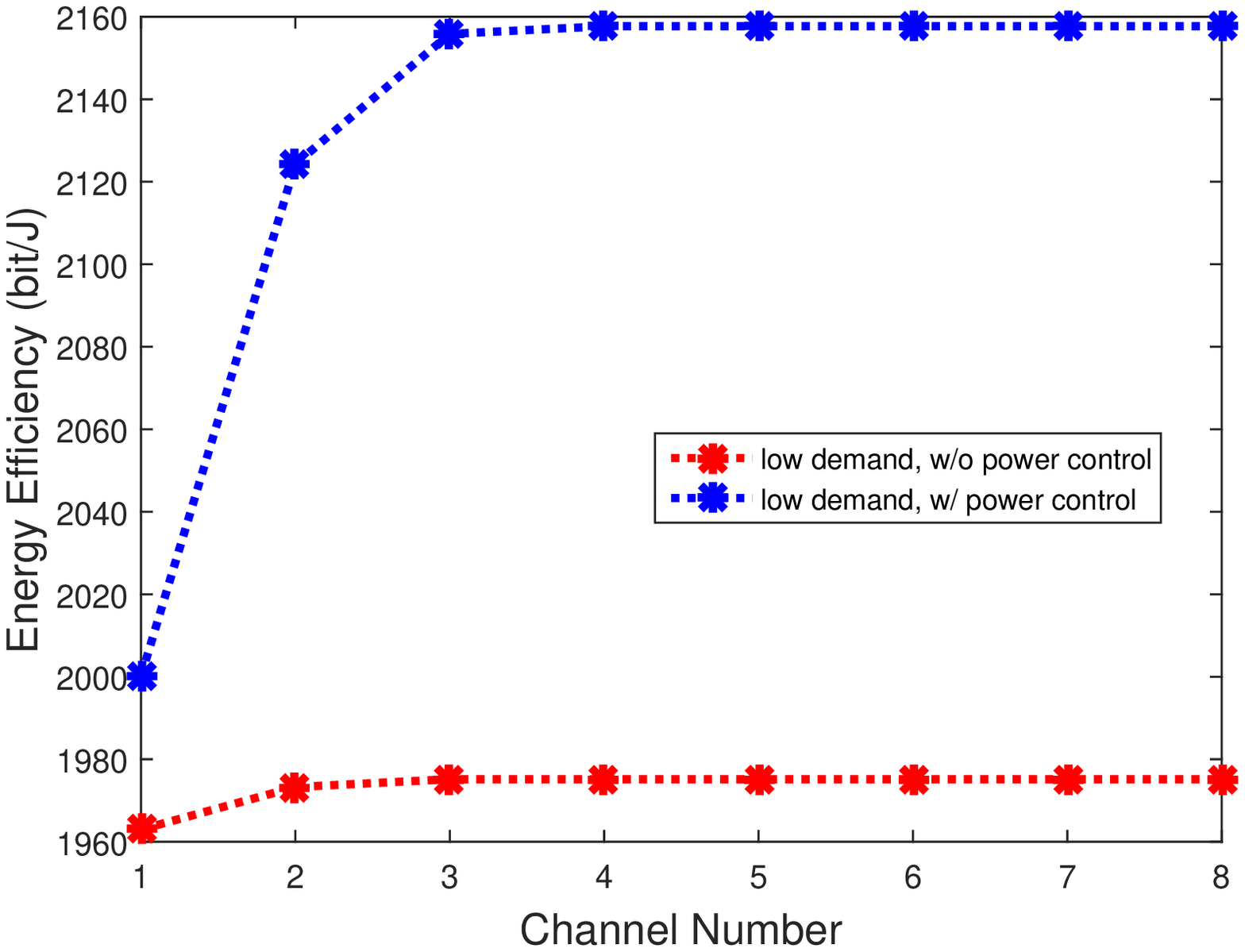}
        }
        \subfigure[2 radio per node.]{
            \includegraphics*[width=2.2in]{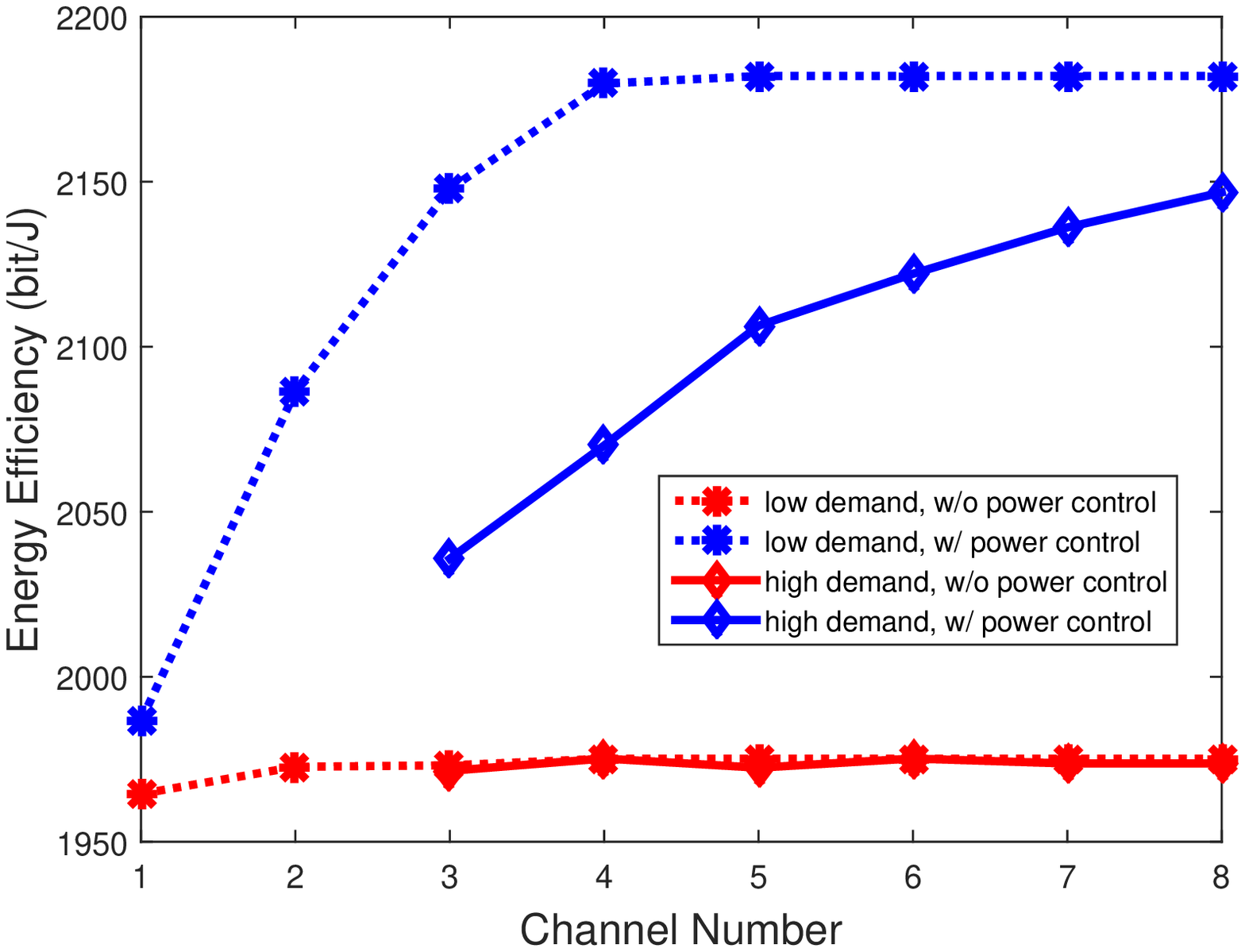}
        }
	    \subfigure[3 radio per node.]{
            \includegraphics*[width=2.2in]{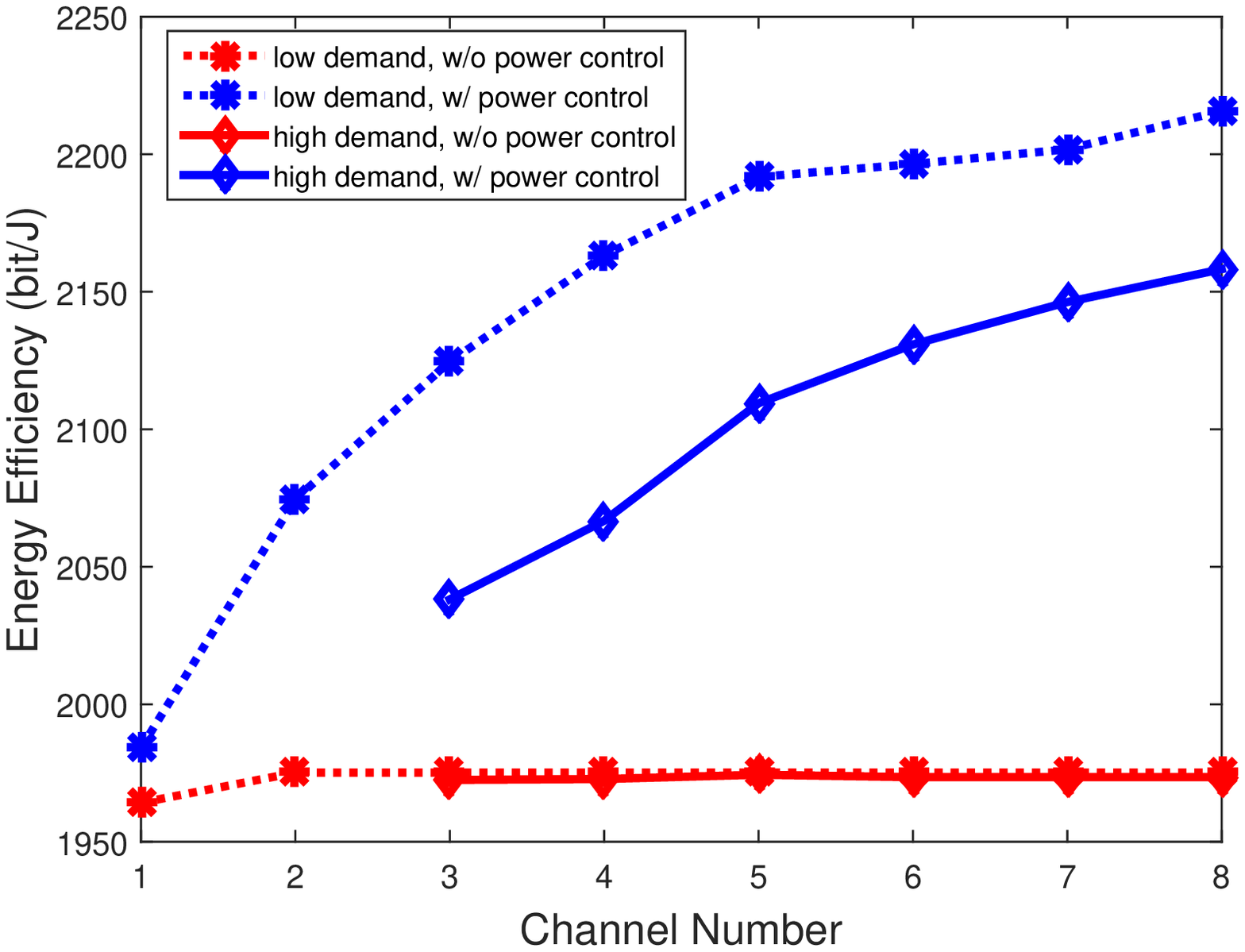}
        }
\caption{Energy efficiency comparison under different network parameters.}\label{fig:radio}
\end{figure*}

The energy efficiency comparison under various numbers of radios and channels is shown in Fig. \ref{fig:radio}. The missing data points (e.g. there is no curve for high demand in Fig. \ref{fig:radio}(a)) is due to no feasible solution found after a large number of rounds (which likely means the original problem is infeasible).

In consistency with the previous subsection, it is observed that the result with power control can outperform that of fixed power (without power control) in all scenarios. As aforementioned, more choices of power levels enable transmissions with lower power when traffic demand is low, thus improving the network energy efficiency.

In Fig. \ref{fig:radio}, higher traffic demand often leads to lower energy efficiency. Generally,  higher traffic demand requires more simultaneous transmissions or higher transmit power, which leads to more co-channel interference and degradation of transmission quality as well as energy efficiency. Thus, a direct remedy can be exploiting more channel resources. As seen from Fig. \ref{fig:radio}, when more channels become available, the energy efficiency in the high demand case is not much less than that of low demand.

When there is only one or a few channels in the network, it is more likely that simultaneous transmissions will take place in the same channel and suffer co-channel interference, which will impact energy efficiency. While more channels means transmissions can be separated to different bands and avoid co-channel interference. However, more channels may not lead to better performance all the time, as can be seen in Fig. \ref{fig:radio}. When the number of channels is very large but the number of radios is limited, the extra spectrum resource cannot be fully utilized due to radio conflict. In other words, there are not enough radios to occupy these channel bands. Furthermore, there is no obvious performance improvement from 2 radios to 3 radios when the demand is low, which indicates that the one extra radio can be turned off or put to sleep mode to save energy.

These results may guide choosing proper numbers of radios or channels in the network. If the number of radios is given, choose the least number of channels that can maximize energy efficiency of the network. If the number of channels is given, turn on just enough number of radio interfaces and turn off the extra radios. In addition, the number of radios and channels can be jointly determined according to traffic demand to avoid excessive expenditure of resources.

\subsection{Trade-off between SE and EE}

\begin{figure*}[htbp]
  \centering
        \subfigure[SE with bandwidth.]{
            \includegraphics*[width=2.2in]{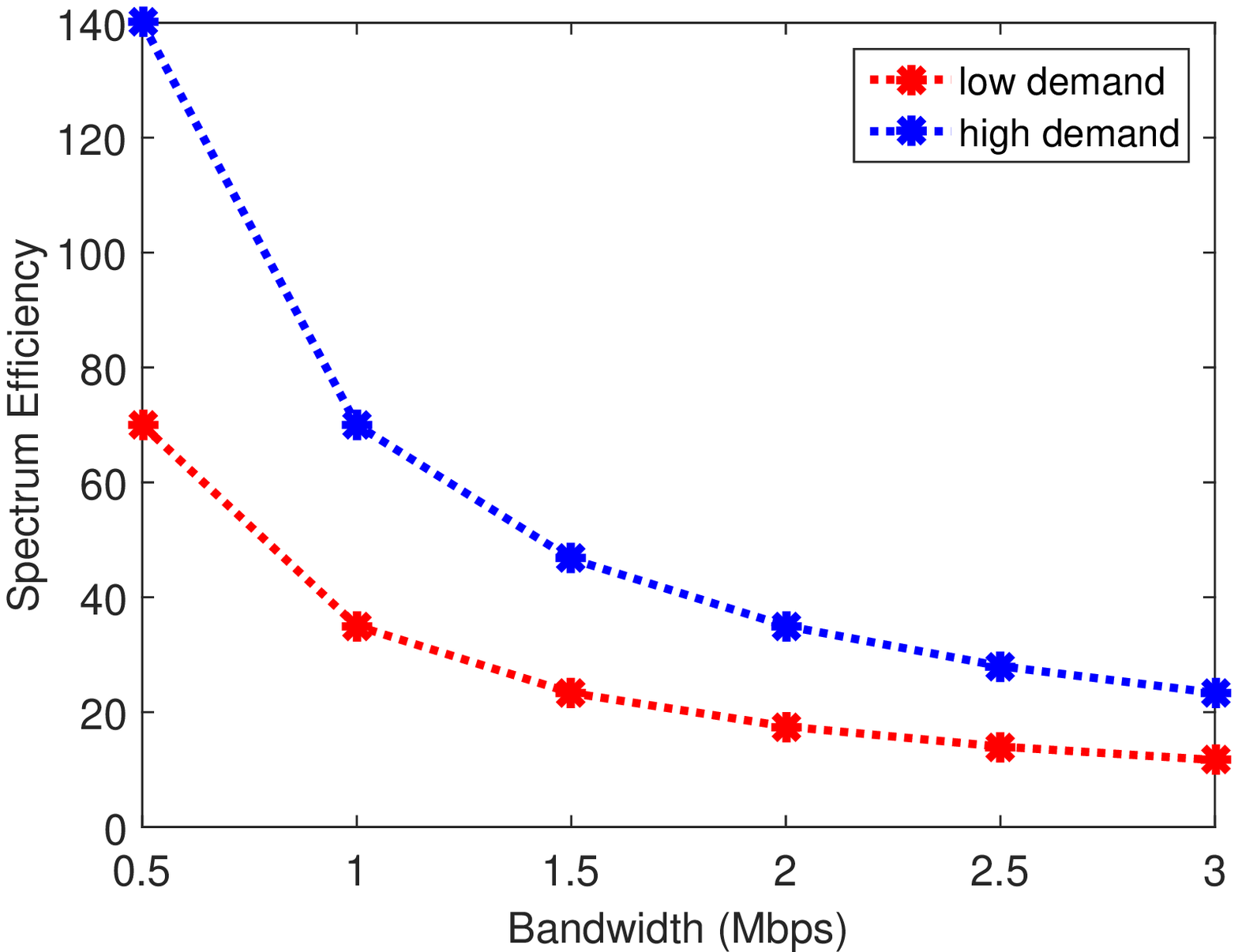}
        }
        \subfigure[EE with bandwidth.]{
            \includegraphics*[width=2.2in]{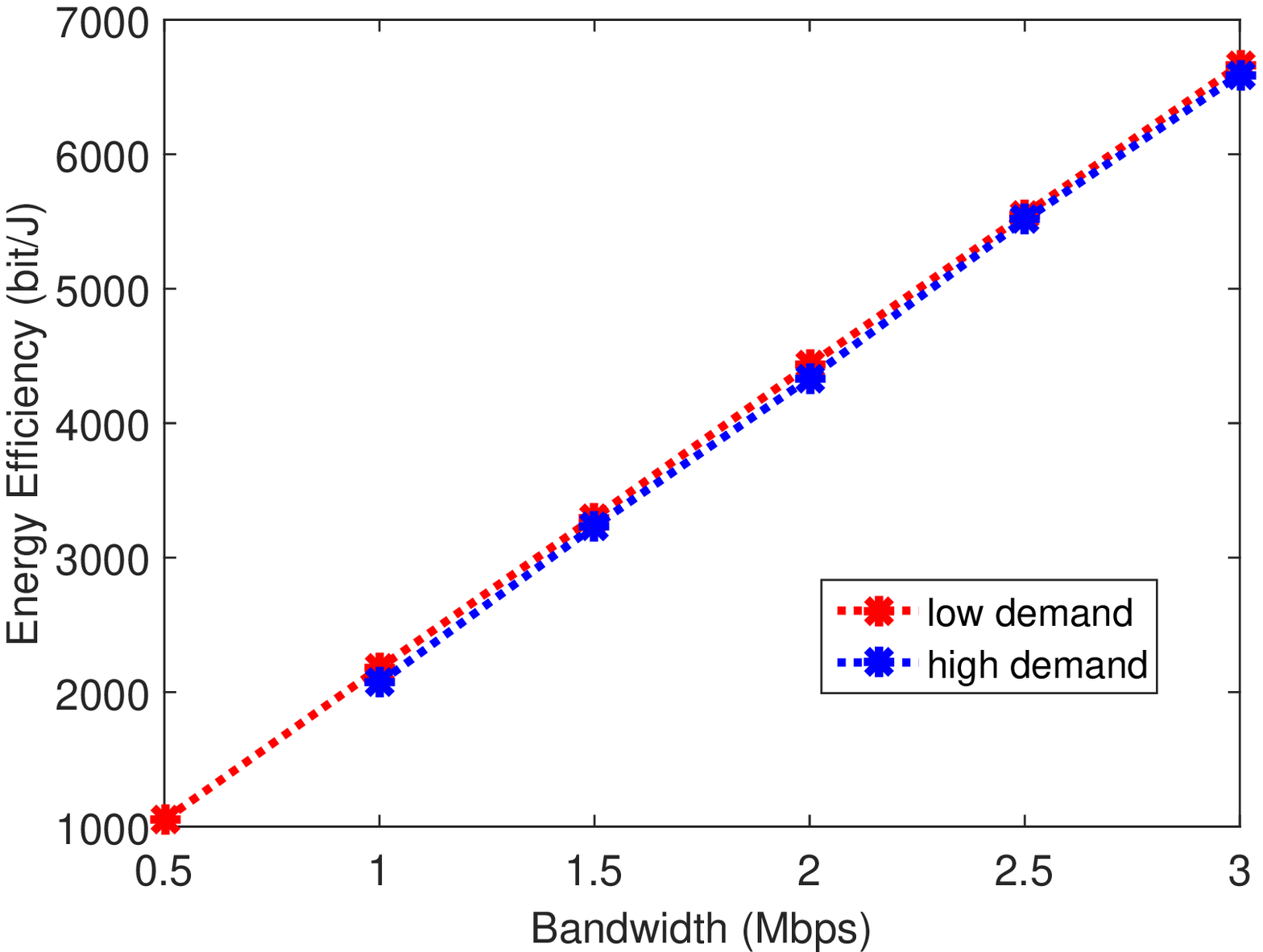}
        }
        \subfigure[SEE with bandwidth.]{
            \includegraphics*[width=2.2in]{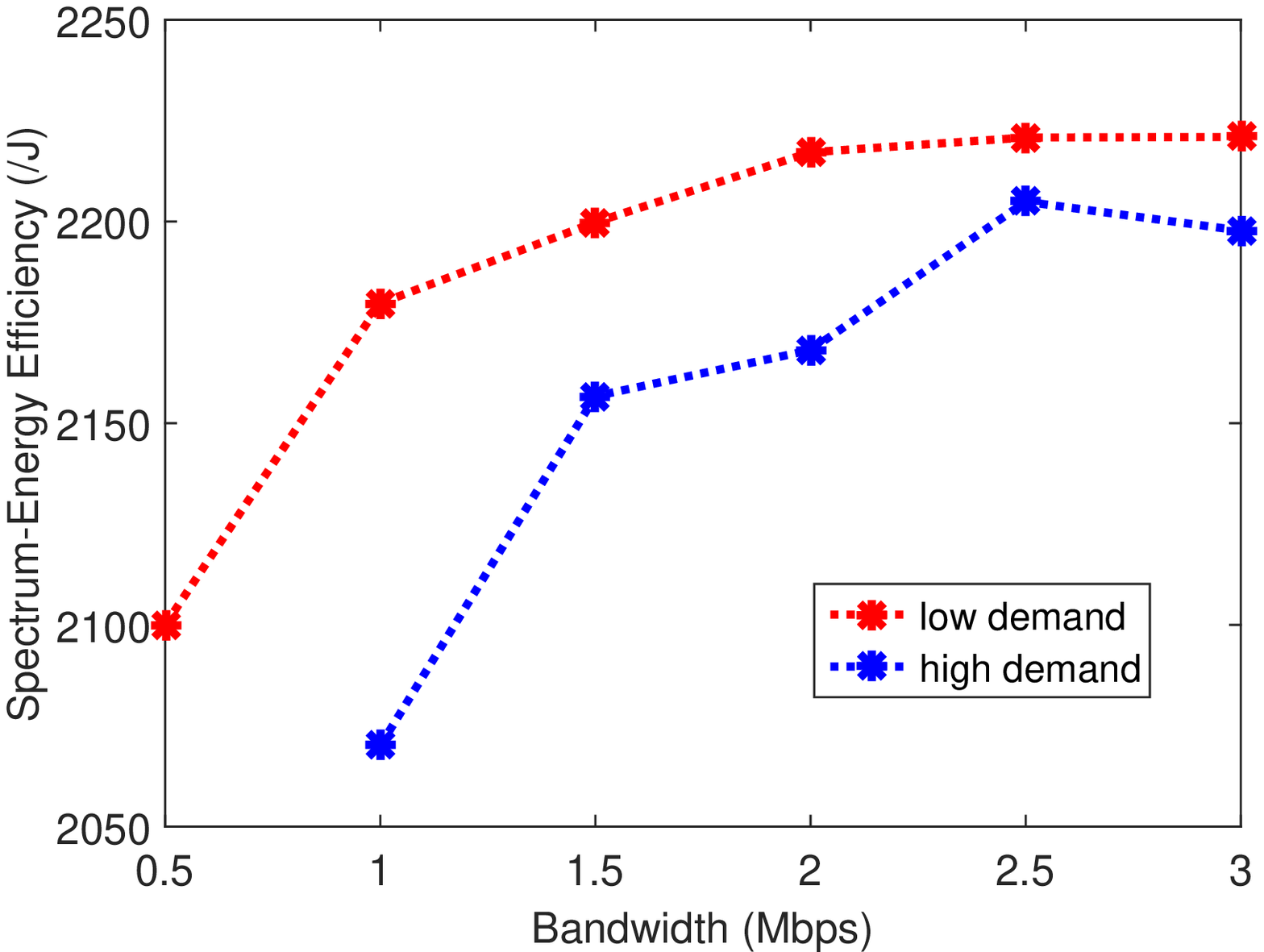}
        }
\caption{SE and EE trade-off under different bandwidths.}\label{fig:SE}
\end{figure*}

Spectrum efficiency (SE) is defined as the ratio of achieved data rate over spectrum resource (i.e., in this paper, the total bandwidth). Generally when EE increases, SE will decrease, and vice versa. Such a trade-off between SE and EE is shown in Fig. \ref{fig:SE}(a)(b). We also compare the corresponding spectrum-energy efficiency (SEE) in Fig. \ref{fig:SE}(c), which is defined as SE divided by energy consumption.

Consider a simple example that if the bandwidth is doubled, then the transmission time of all the links can be halved while the traffic demand can still be satisfied but the energy consumption is halved. In this case, SE is halved and EE is doubled. This simplified analysis can explain the trends in Fig. \ref{fig:SE}(a)(b). However, the effect of bandwidth on EE is more than this. In the example, since the transmission time of all the links is halved, it means less simultaneous transmissions are required, which will potentially reduce co-channel interference and thus further improve energy efficiency. As a result, EE will be increased more than twice. Following this analysis, the trend in Fig. \ref{fig:SE}(c) can be explained that SEE will grow as the bandwidth increases.

\subsection{Computation Time}

The computation time of the decomposition algorithm is determined by the time consumption in each round and the number of rounds to converge, where the latter depends on the network topology and parameter setting. Even if the topology and parameters are the same, due to the randomness introduced from breaking tied values, the required number of rounds may also be different. Since the iterative process can be viewed as finding more accurate evaluation of TPs and finding better TPs based on the evaluation, the larger scale the network is, the more difficult in finding the final result. As a result, generally as the scale of network increases, it will take more rounds to converge as shown in Table \ref{time}.

Within each round, the computation mainly consists of two stages: the master stage of solving an LP problem and the sub-problem stage of a greedy algorithm. Table \ref{time} shows the average computation time per round in the first 100 rounds for these two stages under different network configurations.

The computation time of solving master problem is mainly determined by the size of the constraint matrix $\mathcal{A}^{(k)}$. In our case, the number of rows of $\mathcal{A}^{(k)}$ is equal to the number of links, and its number of columns is initialized also at the number of links but increased by one in each round. From the result we can observe that the computation time of master stage is almost linear with the number of links.

The time consumption in solving sub-problem with greedy algorithm also grows with the number of links. Within the greedy algorithm, links will be gradually activated until the total utility is maximized. The number of activated links or the number of rounds in greedy algorithm also affects the computation time. For example, in Table \ref{time}, the 3rd and 4th cases are both with 560 links, but the average time consumed in greedy algorithm for the 4th case is longer than that of the 3rd case. This is because in the 2-radio case, more links can be activated simultaneously and correspondingly the greedy algorithm will run more rounds to add these links.

\begin{table}[!t]
\renewcommand{\arraystretch}{1.3}
\caption{Computation Time}
\label{time}
\centering
\begin{tabular}{c|c|c|c}
\hline
\hline
configuration & rounds & master-time & sub-time\\
\hline
 1-2-140& $<100$ &  0.03s & 0.03s \\
 1-5-350& $150\sim 200$ & 0.07s & 0.08s \\
 1-8-560& $200\sim 250$ & 0.12s & 0.14s \\
 2-2-560& $50\sim 100$ & 0.10s & 0.16s \\
 2-5-1400& $200\sim 300$ & 0.27s & 0.59s \\
 2-8-2240& $400\sim 500$ & 0.60s & 1.39s \\
 3-2-1260& $200\sim 300$ & 0.21s & 0.47s \\
 3-5-3150& $>500$ & 0.69s & 2.01s \\
 3-8-5040& $>1000$ & 1.32s & 4.67s \\
\hline
\hline
\end{tabular}\\
\vspace{1mm}
*(1-2-140 means 1 radio per node, totally 2 channels and 140 links)
\end{table}

\section{Conclusion}

In this paper we have investigated energy-efficient joint resource allocation in generic wireless networks with multi-dimensional resource space. We have formulated a joint scheduling and power control problem which aims at minimizing energy consumption of the network while satisfying flow demand requirements. The large-scale problem with coupled variables has been solved efficiently by decomposition based on DCG method and greedy algorithm. The solution provides a joint allocation of radio, channel, transmit power as well as scheduling and routing. Theoretical analysis on the performance of the proposed algorithm has been conducted, and numerical results demonstrated that the proposed algorithm can improve energy efficiency of MR-MC networks with joint scheduling and power control.

\bibliographystyle{IEEEtran}

\end{document}